\documentclass[aps,a4paper,showpacs,preprintnumbers,amsmath,amssymb,
  floatfix]{revtex4}
\usepackage{dcolumn}
\usepackage{bm}
\usepackage{amsmath,amssymb}
\usepackage{booktabs,subfigure}
\usepackage{graphicx,psfrag}
\usepackage{epsfig}
\usepackage{color} 
\usepackage{rotating}
\usepackage{axodraw}
\usepackage{slashed}
\usepackage{geometry}
\usepackage{rotating}
\usepackage{subfigure}
\usepackage{caption}
\usepackage{pstricks}
\usepackage{color}
\allowdisplaybreaks[3]
\newcommand{\be}{\begin{equation}}
\newcommand{\ee}{\end{equation}}
\newcommand{\bea}{\begin{eqnarray}}
\newcommand{\eea}{\end{eqnarray}}
\newcommand{\cw}[1][{}]{\ensuremath{\cos^{#1} \theta_{W}}}
\newcommand{\sw}[1][{}]{\ensuremath{\sin^{#1} \theta_{W}}}
\newcommand{\tw}[1][{}]{\ensuremath{\tan^{#1} \theta_{W}}}

\newcommand{\M}{\ensuremath{\mathcal{M}}}

\numberwithin{equation}{section} 
\usepackage{epsfig}
\usepackage{dcolumn} % Align table columns on decimal point
\usepackage{bm} % bold math

\def\gsim{\lower0.5ex\hbox{$\:\buildrel >\over\sim\:$}}
\def\lsim{\lower0.5ex\hbox{$\:\buildrel <\over\sim\:$}}

\begin{document}

\title{
Radiative Production of Lightest Neutralinos in $e^+ e^-$ collisions 
in Supersymmetric Grand Unified Models }

\vskip 2cm

\author{P. N. Pandita,$^1$
              %\footnote{\tt{Electronic address: ppandita@nehu.ac.in}}%
~Monalisa Patra$^2$}
              %\footnote{\tt{Electronic address: monalisa@cts.iisc.ernet.in}}}

\affiliation{$^1$ Department of Physics, North Eastern Hill University, 
                  Shillong 793 002, India \\ %
             $^2$ Centre for High Energy Physics, 
                  Indian Institute of Science, Bangalore 560 012, India}
%%%%

%\date{}
\thispagestyle{myheadings}

%\maketitle

\vskip 5cm

\begin{abstract}
\noindent
We study the  production of the lightest neutralinos 
in the radiative process $e^+e^- \to \tilde\chi^0_1 \tilde\chi^0_1\gamma$ 
in supersymmetric  models with grand unification.
We consider models wherein the standard model gauge group 
$SU(3)_c \times SU(2)_L \times U(1)_Y$ 
is unified into the grand unified gauge groups $SU(5)$, or $SO(10)$.  
We study this process  at energies that may be accessible at 
a future International Linear Collider.  
We  compare and contrast the dependence of the signal cross
section on the  grand unified gauge group, and different representations
of the grand unified gauge group,
into which the standard model gauge group is unified.
We carry out  a comprehensive study of the radiative production process 
which includes higher order  QED corrections  in our calculations.
In addition we carry out a detailed study 
of the  background to the signal process 
coming from the
Standard Model radiative neutrino production  
$e^+e^- \to \nu \bar\nu \gamma$, as well
as from the radiative production of the scalar partners of the 
neutrinos~(sneutrinos) $e^+e^- \to \tilde\nu \tilde\nu^\ast \gamma$. 
The latter can be  a major supersymmetric background to the radiative 
production of neutralinos when the sneutrinos decay invisibly.  It is
likely that the radiative production of the lightest 
neutralinos may be  a viable channel to study supersymmetric partners of the 
Standard Model particles at the
first stage of a International Linear Collider, where  heavier sparticles 
may be too heavy to be produced in pairs.

\end{abstract}
\maketitle

%%%%%%%%%%%%%%%%%%%%%%%%%%%%%%%%%%%%%%%%%%%%%%%%%%%%%%%%%%%%%%%%%%%%%%%%%%%%%
\section{Introduction}
\label{sec:intro}
Supersymmetry~(SUSY)~\cite{Wess:1992cp} is a leading candidate
for physics beyond the  standard model~(SM). In supersymmetric models
the Higgs sector of the SM is  technically 
natural~\cite{'tHooft:1980xb}. Since supersymmetry is
not an exact symmetry in nature,  it must be broken  
in realistic models of  supersymmetry.
Although the precise manner in which SUSY is broken is not known  at
present, in actual practice  necessary SUSY breaking can be introduced 
through soft supersymmetry breaking terms that do not reintroduce quadratic 
divergences in the Higgs mass, and thereby do not disturb the stability of 
the hierarchy between the
weak scale and a  large scale~(grand unified~(GUT),  or Planck
scale).  
The simplest implementation of the 
idea of softly broken supersymmetry 
is the Minimal Supersymmetric Standard Model~(MSSM) obtained by  
introducing the supersymmetric partners of the SM states, and introducing an 
additional Higgs doublet, with opposite hypercharge to that of the SM 
Higgs doublet, in order to cancel the gauge 
anomalies and generate masses for all the fermions of the Standard 
Model~\cite{Nilles:1983ge}. If we want 
broken supersymmetry to be effective in protecting the Higgs mass
against large radiative corrections, then 
the supersymmetric partners of the Standard Model~(SM) particles cannot be much 
heavier than about $1$~TeV.  The  discovery of the superpartners 
of the SM particles
is one of the main goals of present and future accelerators.

When the SM gauge symmetry $SU(2) \times U(1)$ is broken, 
the fermionic partners of the two Higgs doublets~($H_1, H_2$) 
of the MSSM mix with the fermionic partners
of the gauge bosons, resulting in  four neutralino states $\tilde \chi^0_i$,
$i = 1, 2, 3, 4$, and two chargino states  $\tilde \chi^{\pm}_j$,
$j = 1, 2.$  
In several  models of low energy supersymmetry, the lightest neutralino
is typically the lightest supersymmetric particle~(LSP).
In the MSSM, if we assume $R$-parity~($R_p$) conservation, then
the lightest supersymmetric particle is absolutely stable. 
There has been extensive study of the 
neutralino states of the minimal supersymmetric
standard model, and its 
extensions~\cite{Bartl:1989ms,Bartl:1986hp,Pandita:1994ms,Pandita:1994vw,
Pandita:1994zp,Pandita:1997zt,
Choi:2001ww,Huitu:2003ci,Huitu:2010me}, 
because the lightest neutralino, being the LSP, is the end product 
of any process that involves supersymmetric particles in the final state.
In this work we will assume
that the LSP is the lightest neutralino, it is stable, and that it escapes the
collider experiments undetected.  The composition
and mass of the neutralinos and charginos will play a key role in the
search for supersymmetry at high energy accelerators.  
The composition and mass of neutralinos
will also  determine the time-scale of their decays.  The 
implications of mass patterns 
of the neutralinos in models with different particle content, or with 
specific SUSY
breaking patterns were considered in some detail 
in~\cite{Pandita:1994zp,Huitu:2003ci, Huitu:2010me}.

%____________________________________________________________________

At present an indirect phenomenological evidence for supersymmetry is 
obtained from the unification of the gauge couplings of the 
Standard Model in supersymmetric grand unified 
theories~(GUTS)~\cite{Dimopoulos:1981yj, Langacker:1995fk}. 
Furthermore, one of the most important prediction of grand unification 
is that of baryon number violating interactions, leading to proton decay.
However, in supersymmetric grand unified theories proton decay is
much slower than in nonsupersymmetric grand unified theories.
The reason for this is that the unification scale in supersymmetric GUTS
is of the order of $M_{\rm GUTS} \sim {\rm few} 10^{16}$ GeV, which is
about $20-30$ times larger than the corresponding scale in non-SUSY GUTS.
Thus, proton decay via gauge boson exchange is negligible and 
main decay arises from dimension-$5$ operators with higgsino exchange.
This leads to a rate of proton decay which is close to
the  existing bounds~\cite{AFM}. Furthermore, the range of neutrino 
masses as indicated by current experiments, when interpreted in terms of 
see-saw mechanism, point towards a large scale consistent with
the scale $M_{\rm GUTS}$ of supersymmetric grand unification~\cite{alfe}.     

It is, thus, natural to study the phenomenology of neutralino 
and charginos in an underlying grand unified theory. 
Most of the studies involving neutralinos
and charginos in the minimal supersymmetric standard model
have been performed with universal gaugino masses at the
grand unification scale~\cite{Basu:2007ys}. 
The neutralino and chargino masses depend on the 
soft $SU(2)_L$  and $U(1)_Y$  gaugino masses $M_2$ and $M_1$,
the higgs(ino) parameter $\mu$, and $\tan\beta \equiv v_2/v_1$,
where $v_2$ and $v_1$ are the vacuum expectation values of the 
neutral components of the two Higgs doublets $H_2$ and $H_1$.  
Most of the models assume the gaugino mass universality at the
GUT scale, i.e. $M_1 = M_2 = M_3$, where $M_3$ is the 
$SU(3)_C$ soft gaugino mass.
However, there is no specific  theoretical reason 
for the choice of universal masses at the grand unification scale.  
It is possible to have nonuniversal gaugino masses
in grand unified theories wherein the 
standard model gauge group is embedded in a grand unified
gauge group.  In a given supersymmetric model
gaugino masses are generated from higher dimensional
interaction terms involving  gauginos and auxiliary parts of chiral
superfields~\cite{Cremmer:1982wb}.  
As an example, in $SU(5)$
grand unified theory~(GUT), the auxiliary part of a chiral
superfield in  higher dimensional terms can be in the
representation {\bf 1}, {\bf 24}, {\bf 75}, or {\bf 200}, or, in general,
some combination of these representations. When the
auxiliary field of one of the $SU(5)$ nonsinglet chiral superfields
obtains a vacuum expectation value (VEV), then the resulting
gaugino masses are nonuniversal at the grand unification scale. 
Similar conclusions
hold for other supersymmetric grand unified models.
Furthermore, nonuniversal
supersymmetry breaking masses are a
generic  feature in some of the  realistic supersymmetric models.
For example, in anomaly mediated supersymmetry breaking models  
the gaugino masses are not unified \cite{Randall:1998uk, Huitu:2002fg}, 
and hence are not universal.

From above discussion it is clear, that 
the phenomenology of supersymmetric models depends
crucially on the composition of neutralinos and charginos.  
This in turn depends on the soft gaugino mass parameters
$M_2$ and $M_1$, besides the parameters $\mu$ and $\tan\beta$.
Since most of the models discussed in the literature
assume gaugino  mass universality at the GUT scale,
it is important to investigate the changes in the phenomenology
of broken supersymmetry which results from the 
changes in the composition of neutralinos
and charginos that may arise because of the changes in the pattern
of soft gaugino masses  at the grand unification scale~\cite{Ellis:1985jn}. 
The consequences of nonuniversal gaugino masses at the grand unified scale 
and the resulting  change in boundary conditions 
has been considered in several papers.
This includes the  study of constraints arising from different
experimental measurements~\cite{Barger:1998hp,Huitu:1999vx,Djouadi:2001fa},
and in the study  of
supersymmetric dark matter candidates~\cite{Bertin:2002sq,Corsetti:2000yq}.  
 
One of the major goals of high energy   colliders is to 
discover the supersymmetric partners of the Standard Model particles.  
In particular, a high energy   $e^+ e^-$ linear collider
with a  center-of-mass
energy of $\sqrt s = 500$ GeV in the first stage, and 
with a high luminosity ${\mathcal L}=500~ {\rm fb}^{-1}$,  
will be  important  in determining the parameters of the
broken supersymmetric model with a high  
precision~\cite{Aguilar-Saavedra:2001rg,Abe:2001nn,Abe:2001gc, 
Weiglein:2004hn,Aguilar-Saavedra:2005pw}.  

Recently in~\cite{Basu:2007ys, Pandita:2011eh}
a detailed study of the radiative 
production of neutralinos in electron-positron collisions
in low energy supersymmetric models with
universal gaugino masses at the grand unified scale 
has been carried out.
In this paper we shall carry out a study of the 
implications of the nonuniversal
gaugino masses, as they arise in grand unified theories,
for the production of lightest neutralinos in 
electron-positron collisions. 
Since in a large class of models of supersymmetry 
the  lightest neutralino is expected to be the
lightest supersymmetric particle, it is one of the first states 
to be produced at the colliders.  
At an electron-positron collider, such as the  
International Linear Collider~(ILC), the lightest neutralino
can be directly produced 
in pairs~\cite{Bartl:1986hp,Ellis:1983er}.
In collider experiments it will escape
detection such that the direct production of the lightest
neutralino pair is invisible.

One must, therefore, look for the signature of 
neutralinos in electron-positron colliders 
in the radiative production process
\begin{equation}
e^+ + e^-\to\tilde\chi_1^0 + \tilde\chi_1^0 + \gamma.
\label{radiative}
\end{equation}
The signature of this process is  a single high energy photon with 
missing energy  carried away by the neutralinos. 
In this paper we carry out  a detailed study
of  the process (\ref{radiative}) in supersymmetric
grand unified theories with nonuniversal boundary conditions
at the grand unified scale. 
Note that this process is suppressed by the square of the 
electromagnetic coupling. However,  it might be the
first process where the lightest supersymmetric particles
could  be observed at $e^+ e^-$ colliders.  
The process~(\ref{radiative}) has been
studied in detail in the minimal supersymmetric  
model ~\cite{Fayet:1982ky,Ellis:1982zz, Grassie:1983kq,
Kobayashi:1984wu,Ware:1984kq,Bento:1985in, Chen:1987ux,Kon:1987gi,
Choi:1999bs, Datta:1996ur, Ambrosanio:1995it}, in  various approximations. 
Calculations have also 
been carried out for MSSM using general neutralino 
mixing~\cite{ Choi:1999bs, Datta:1996ur, Ambrosanio:1995it}.
This process has also been studied in detail in the 
next-to-minimal supersymmetric model~(NMSSM)~\cite{Basu:2007ys, Pandita:2011eh}.
On the other hand different LEP collaborations~\cite{Heister:2002ut, 
Abdallah:2003np, Achard:2003tx, Abbiendi:2002vz,Abbiendi:2000hh}
have studied the  signature
of radiative neutralino production in detail, but have found
no deviations from the SM prediction. Thus, 
they have only been able to set  bounds 
on the masses of supersymmetric particles~\cite{Heister:2002ut,
Abdallah:2003np,Achard:2003tx,Abbiendi:2000hh}.  

We recall that
in the SM the radiative neutrino process
\begin{equation}
e^+e^- \to  \nu + \bar\nu + \gamma, 
\label{radiativenu} 
\end{equation}
is the leading process with the same signature as (\ref{radiative}). The 
cross section  for the process  (\ref{radiativenu})
depends on the number $N_\nu$ of light neutrino
species~\cite{Gaemers:1978fe}.  This process acts as a main background
to the radiative neutralino production process  (\ref{radiative}).
There is also a  supersymmetric background  to the process (\ref{radiative})
coming from radiative sneutrino production  
\begin{equation}
e^+e^- \to \tilde\nu + \tilde\nu^\ast + \gamma.
\label{radiativesnu}
\end{equation}
We shall consider both these processes as well, since they form the 
main background to the  radiative process (\ref{radiative}).

%_____________________________________________
The layout of the paper is as follows. 
In Sec.~\ref{sec:gaugino mass patterns}, we review  different patterns
of gaugino masses that arise in grand unified theories. We will consider
grand unified theories based on $SU(5)$ and $SO(10)$ gauge groups,
and discuss the origin of nonuniversal gaugino masses in 
these models. Here we also calculate the elements of the mixing matrix
which are relevant for obtaining the couplings of the lightest
neutralino to the electron, selectron, and $Z$ boson
which control  the radiative neutralino 
production process (\ref{radiative}). Furthermore, we  also describe in detail
the typical set of input parameters that are used in our numerical
evaluation of cross sections. The set of parameters that we use
are obtained by imposing various experimental and theoretical
constraints discussed in Appendix~\ref{exp_cons} on the parameter 
space of the minimal supersymmetric standard
model with underlying grand unification. These constraints 
will also be used throughout  to arrive at the 
allowed parameter space for different models in this paper.

Furthermore, we expect low energy observables from flavor physics
and $g-2$ to have some impact in constraining various models
studied in this paper. The  study of  the impact of
these observables on our analysis is, however,  beyond the scope of
present  paper.

In Sec.~\ref{sec:radiative cross section} we summarize the phase space for 
the signal process, and also the cuts on the photon angle and energy that
are  used to regularise the infrared and collinear divergences in the tree
level cross section. In Sec.~\ref{sec:numerical} we evaluate  
the cross section for the signal process (\ref{radiative}) in 
different grand unified theories with nonuniversal gaugino masses, 
using the set of parameters obtained in 
Sec.~\ref{sec:gaugino mass patterns} for different patterns of gaugino
mass parameters at the grand unified scale.
We have included higher order QED radiative corrections
in our calculations. 
We  also compare and contrast 
the results so obtained  with the 
corresponding cross section in the MSSM with universal gaugino masses
at the grand unified scale.  The dependence of the
cross section on the parameters of the neutralino sector, and on the
selectron masses is also studied  numerically.  

In Sec.~\ref{sec:backgrounds} we discuss the backgrounds  to 
the radiative neutralino production process (\ref{radiative}) from
the SM and supersymmetric processes.  An excess of photons 
from radiative neutralino production over the backgrounds measured
through statistical significance is also discussed here and
calculated for different grand unified
models. 
Our results are summarized and the conclusions presented in 
Sec.~\ref{sec:conclusions}.

%%%%%%%%%%%%%%%%%%%%%%%%%%%%%%%%%%%%%%%%%%%%%%%%%%%%%%%%%%
\section{ Gaugino Mass Patterns in Grand Unified Theories}
\label{sec:gaugino mass patterns}

In this section we shall discuss soft supersymmetry breaking
gaugino mass patterns that arise in $SU(5)$ and $SO(10)$ supersymmetric 
grand unified models,
and the implications of these mass patterns for the
neutralino masses and couplings. In Appendix~\ref{neut mass mat}, we 
summarize our notations for the neutralino mass matrix and 
the couplings of the lightest neutralino that are
relevant to our study\cite{Haber:1984rc, Basu:2007ys}. 
Furthermore, in Appendix~\ref{exp_cons} 
we summarize the current experimental constraints~\cite{lep-chargino,
Yao:2006px, Abdallah:2003xe, Dreiner:2009ic} on the parameters 
of the neutralino mass matrix that we use in our calculations.

\subsection{Universal Gaugino Masses in  Grand Unified Theories}
\label{subsec:universal-gaugino-masses} 
In  supersymmetric models, with gravity 
mediated supersymmetry breaking,
usually denoted as mSUGRA, the soft supersymmetry 
breaking gaugino mass parameters 
$M_1, M_2$, and $M_3$ and the respective gauge couplings $g_i$ 
satisfy the renormalization group
equations~(RGEs)~($|M_3| \equiv  M_{\tilde g}$, the gluino mass) 
\bea
16\pi^2\frac{dM_i}{dt} & = & 2 b_i M_i g_i^2, ~~~~b_i =
\left(\frac{33}{5}, 1, -3\right), \label{gaugino1}\\
16\pi^2\frac{dg_i}{dt} & = & b_i g_i^3 \label{gauge1}
\eea
at the one-loop  order, where $ i = 1, 2, 3 $ refer to the 
$U(1)_Y, SU(2)_L$ and the $SU(3)$ gauge groups, respectively. 
Here, $g_1 =\frac{5}{3}g',\; g_2 = g$, with $g'$ and $g$ as 
$U(1)_Y$, and $SU(2)_L$ gauge couplings, respectively, and $g_3$ 
is the $SU(3)_C$ gauge coupling. With the universal boundary 
conditions on the gaugino masses~($\alpha_i = g_i^2/4\pi, \, i = 1, 2, 3$),
we have
\bea
M_1 & = &  M_2 = M_3 = m_{1/2}, \label{gauginogut}\\
\alpha_1 & = & \alpha_2 =  \alpha_3 = \alpha_G,  \label{gaugegut}
\eea
at the GUT scale $M_G$. Then the  RGEs (\ref{gaugino1}) and (\ref{gauge1}) 
imply that
the soft  gaugino masses scale like gauge couplings:
\bea
\frac{M_1(M_Z)}{\alpha_1(M_Z)} & = & \frac{M_2(M_Z)}{\alpha_2(M_Z)}
= \frac{M_3(M_Z)}{\alpha_3(M_Z)} =  \frac{m_{1/2}} {\alpha_G}.
\label{gaugino2}
\eea
The relation (\ref{gaugino2}) implies that out of three gaugino mass parameters
only one is independent, which we are free to choose as the gluino 
mass $M_{\tilde g}$. The remaining soft gaugino mass parameters can be determined through  
\bea M_1(M_Z) & =
& \frac{5 \alpha}{3 \alpha_3~\cos^2\theta_W}~M_{\tilde g} ~~\simeq~~
0.14~M_{\tilde g},
\label{m3relation}\\
M_2(M_Z) & = & \frac{\alpha}{\alpha_3~\sin^2\theta_W}~M_{\tilde g}
~~\simeq~~ 0.28~M_{\tilde g} ,
\label{m2relation}
\eea where  we have  used the values of various couplings at the $Z^0$
to be
\bea \alpha^{-1}(M_Z) = 127.9, ~~~~~ \sin^2\theta_W = 0.23, ~~~~~
\alpha_3(M_Z) = 0.12.  \eea 
For the gaugino mass parameters this leads to the ratio
\begin{equation}
M_1 : M_2 : M_3 \simeq 1 : 2 : 7.1.
\label{msugra0}
\end{equation} 
The gaugino mass parameters described above are the running masses
evaluated at the electroweak scale $M_Z$. 

Using  the  ratio (\ref{msugra0}) and the constraint (\ref{limits1}), we have
the lower bound  on the parameter $M_1$
\bea
M_1 & \gsim & 50~ {\rm GeV}, \label{msugra2}
\eea
in mSUGRA model. We shall implement  this constraint on the  parameter $M_1$
in our calculations.
%~\cite{Huitu:2010me}. 

For the case of universal gaugino masses at the grand unified scale, we shall
use the set of parameters shown in
Table~\ref{parEWSB}.  We shall call this set of parameters the MSSM 
electroweak symmetry breaking scenario~(EWSB)~\cite{Pukhov:2004ca}.
This scenario has the advantage that it allows us to study 
the dependence of the neutralino masses
and the radiative neutralino production cross section on $\mu$ and $M_2$,
and on the selectron masses.  

\begin{figure}[ht]
\begin{minipage}[b]{0.45\linewidth}
\vspace*{0.65cm}
\centering
\includegraphics[width=6.5cm, height=5cm]{m1m2.eps}
\caption{Dependence of the mass parameters $M_1$ and $M_2$ on the gluino mass $M_{\tilde{g}}$.
Each triangle in the plot refers to a gluino of a particular mass, starting from 1400 GeV,
and successively increasing by 100 GeV.}
\label{fig:m1m2}
\end{minipage}
\hspace{0.7cm}
\begin{minipage}[b]{0.45\linewidth}
\centering
\includegraphics[width=6.5cm, height=5cm]{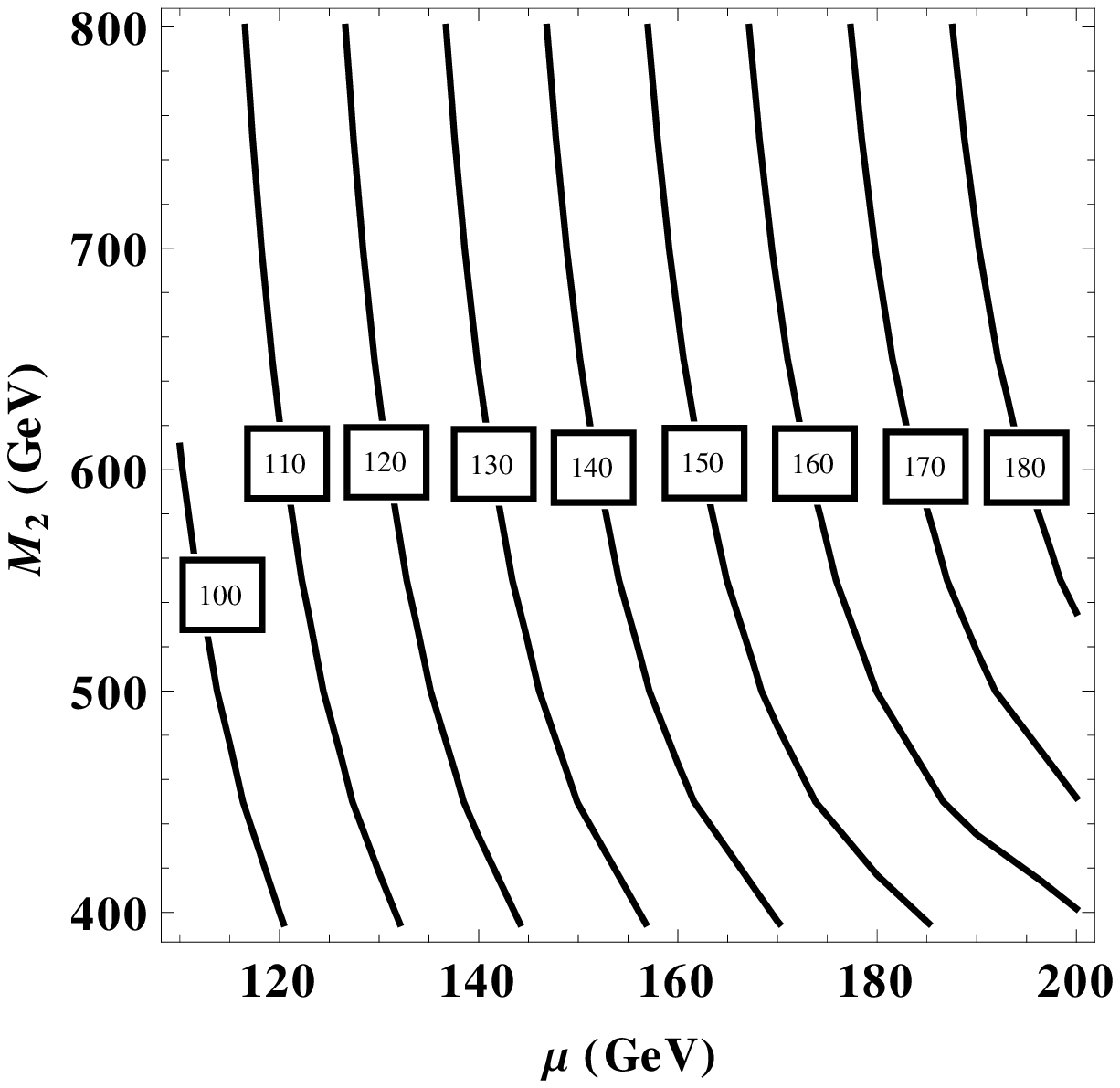}
\caption{The lightest neutralino mass in $\mu - M_2$ plane in case of the MSSM EWSB scenario. The value of $\mu$ 
giving $m_{\tilde {\chi}_1^0} \approx$ 108 GeV, for the smallest value of $M_2$ satisfying the gluino mass constraint, is 
chosen for our analyses. }
\label{fig:mum2ewsb}
\end{minipage}
\end{figure}

The values of different parameters in Table~\ref{parEWSB} 
have been arrived as follows. We first choose the smallest  value of 
$M_3$ to be around $1400$ GeV
as dictated by the experimental constraints on the gluino mass. We then vary it
from $1400$ to $3000$ GeV in steps of $100$ GeV. For the smallest value
of $M_3 = 1400$ GeV, we determine the values of soft 
gaugino mass parameters $M_1$ and $M_2$. Increasing the value of 
$M_3$ in steps of $100$ GeV, we obtain the corresponding values of 
$M_1$ and $M_2$ for different vales of $M_3$. This is shown
in Fig.~\ref{fig:m1m2} as the curve labelled MSSM EWSB.
Then we scan the values of the parameter $M_2$, corresponding to different
values of $M_3$, with values of the parameter $\mu$ 
varying between $110$ to  $200$ GeV such that
the mass of the lightest neutralino lies between
$100$ to $200$ GeV. This is shown as contour plot
in Fig.~\ref{fig:mum2ewsb}. For Table~\ref{parEWSB}
for the MSSM EWSB,
we have chosen the values of $M_2$ and $\mu$ in
Fig.~\ref{fig:mum2ewsb} which correspond
to the lightest neutralino mass of $108$ GeV
and chargino mass larger than $110$ GeV.
Other values can be obtained  by 
choosing a higher mass for the lightest neutralino. 

After selecting the  values of the parameters for our analysis,
we vary the value of $\tan\beta$ in the range so that the top and bottom
quark Yukawa couplings remain perturbative upto the grand unified scale.
For this variation in the value of  $\tan\beta$, we find that 
the mass of the lightest neutralino varies by only $\pm$ (2 $-$ 3)~GeV
compared to its value for  $\tan\beta$ = 10,  with very little change
in the cross section for the radiative neutralino production.
Since our analysis is insensitive to the value of $\tan\beta$,
we have chosen  $\tan\beta = 10$ for definiteness.

We note from the Appendix~\ref{neut mass mat}, the couplings of the
lightest neutralino to electrons, selectrons, and $Z$ bosons, which are 
used for the calculation of the radiative neutralino production
cross section, are determined by the corresponding elements of the
neutralino mixing matrix $N_{ij}$.  For the MSSM EWSB scenario of 
Table~\ref{parEWSB} the composition of the lightest neutralino is given by
\begin{eqnarray}
N_{1j} & = & (0.345, -0.175, 0.703, -0.596).
\label{ewsbcomp}
\end{eqnarray}
The lightest neutralino in this scenario is dominantly a higgsino.

From Table~\ref{feynmandiag} in Appendix~\ref{neut mass mat}, we see that 
for the composition of the neutralino in (\ref{ewsbcomp}), 
the neutralino - $Z^0$ coupling  is enhanced compared to the
coupling of the lightest neutralino with right and left selectrons 
$\tilde e_{R,L}$.

%\medskip

%
%%%%% Table MSSM EWSB Scenario %%%%%%%%
%
\begin{table}%[b]
\renewcommand{\arraystretch}{1.0}
\begin{center}
\vspace{0.5cm}
\begin{tabular}{|c|c|c|c|}
\hline
$\tan\beta$ = 10 &$\mu$ = 130 GeV &$M_1$ = 197 GeV &$M_2$ = 395 GeV \\
\hline
$M_3$ = 1402 GeV    &$A_t$= 2300 GeV &$A_b$= 2300 GeV &$A_{\tau}$= 1000 GeV \\
\hline 
$m_{\chi_1^0}$ = 108 GeV &$m_{\chi_1^\pm}$ = 125 GeV 
 &$m_{\tilde e_R}$ = 156.2 GeV &$m_{\tilde\nu_e}$ = 136 GeV \\
%\cline{1-3}
\hline
$m_{\chi_2^0}$ = 140 GeV &$m_{\chi_2^\pm}$ = 422.4 GeV  
 &$m_{\tilde e_L}$ = 156.7 GeV &$m_h$ = 124.3 GeV \\ 
\hline
\end{tabular}
\end{center}
\vspace{-0.5cm}
\caption{Input parameters and resulting masses of various 
states in  MSSM EWSB scenario.}
\renewcommand{\arraystretch}{1.0}
\label{parEWSB}
\end{table}
%%%%%%%%%%%%%%%%%% End table MSSM EWSB scenario %%%%
%

As a benchmark, we  shall use the radiative neutralino
cross section for the MSSM with universal boundary condition~(\ref{gauginogut})
for the gaugino mass parameters at the GUT scale. For this,
we shall work with the set of parameters  as shown in Table~\ref{parEWSB}.
%___________________________________
\subsection{Nonuniversal Gaugino Masses in  Grand Unified Theories}
\label{subsec:non-universal-gaugino-masses}
We now consider the neutralino masses and mixing in the minimal 
supersymmetric standard model with nonuniversal boundary conditions at 
the GUT scale which arise in $SU(5)$ and $SO(10)$ grand unified theories.
The masses and the composition of neutralinos and charginos are
determined by the soft supersymmetry breaking gaugino masses $M_1$
and $M_2$, respectively, the supersymmetric Higgs~(ino) mixing parameter 
$\mu$, and the ratio of the vacuum expectation values of the two
neutral Higgs bosons $H_1^0$ and $H_2^0$, $\langle H_2^0\rangle
/\langle H_1^0\rangle \equiv \tan\beta$. As discussed in 
subsection~\ref{subsec:universal-gaugino-masses}, in the 
simplest supersymmetric
model with universal gaugino masses, $M_i~(i = 1, 2, 3)$ are
taken to be equal at the grand unified scale. However, in
supersymmetric theories with an underlying grand unified gauge group,
the gaugino masses need not necessarily 
be equal at the GUT scale. In this Section
we briefly review the non universality of gaugino masses as it arises in 
$SU(5)$ and $SO(10)$ supersymmetric grand unified theories, and its 
implications for the neutralino masses and couplings.

In grand unified supersymmetric models the soft gaugino masses are
generated by coupling the field strength superfield  $W^a$ to
$f(\Phi)$, where  $f(\Phi)$ is the gauge kinetic function
which is an analytic function of the chiral superfields  $\Phi$ 
in the theory~\cite{Cremmer:1982wb}. 
The Lagrangian for the coupling of gauge kinetic function to the
gauge field strength can be  written as
\bea
{\cal L}_{g.k.} \; & = & \;
\int d^2\theta f_{ab}(\Phi) W^{a}W^{b}
+h.c.,
\label{gk}
\eea
where $a$ and $b$ refer to gauge group indices, and repeated indices are 
summed over.The gauge kinetic function $f_{ab}(\Phi)$ can be written as
\bea
f_{ab}(\Phi) & = & f_0(\Phi^s)\delta_{ab}
+ \sum_n f_n(\Phi^s){\Phi_{ab}^n\over M_P} +
\cdot \cdot \cdot \cdot \cdot \cdot.
\eea
%In grand unified supersymmetric models, non-universal gaugino masses 
%are generated by a non-singlet chiral superfield $\Phi^n$ that appears 
%linearly in the gauge kinetic function 
%$f(\Phi)$, which is an analytic function of the 
%chiral superfields $\Phi$ in the theory \cite{Cremmer:1982wb}.
%The gaugino masses are generated from the coupling of 
%the field strength superfield $W^a$ with $f(\Phi)$,
%when the auxiliary part $F_\Phi$ of a chiral superfield $\Phi$  in 
%$f(\Phi)$ gets a VEV.
%The Lagrangian for the coupling of gauge kinetic function to the 
%gauge field strength can be  written as
%\bea 
%{\cal L}_{g.k.} \; & = & \;
%\int d^2\theta f_{ab}(\Phi) W^{a}W^{b}
%+h.c.,
%\label{gk}
%\eea
%where $a$ and $b$ refer to gauge group indices, and repeated 
%indices are summed 
%over. The gauge kinetic function $f_{ab}(\Phi)$ is given by
%\bea
%f_{ab}(\Phi) & = & f_0(\Phi^s)\delta_{ab} 
%+ \sum_n f_n(\Phi^s){\Phi_{ab}^n\over M_P} + 
%\cdot \cdot \cdot \cdot \cdot \cdot. 
%\eea
Here  $\Phi^s$ and the  $\Phi^n$ denote the singlet and 
the non-singlet chiral superfields, respectively. Also,
$f_0(\Phi^s)$ and $f_n(\Phi^s)$ are functions of gauge singlet
superfields $\Phi^s$, and $M_P$ denotes  some large scale, e.g. 
the Planck scale.  When the auxiliary part $F_\Phi$ of a chiral superfield
$\Phi$ in $f(\Phi)$ gets a VEV $\langle F_\Phi \rangle$, 
the interaction~(\ref{gk}) gives rise to soft gaugino masses:  
\bea
{\cal L}_{g.k.} \; \supset \;
{{{\langle F_\Phi \rangle}_{ab}} \over {M_P}}
\lambda^a \lambda^b +h.c., 
\eea
where $\lambda^{a,b}$ are gaugino fields. Here, we have denoted  
by $\lambda^1$, $\lambda^2$ and $\lambda^3$  as the 
$U(1)$, $SU(2)$ and $SU(3)$ 
gaugino fields, respectively. Since the gauginos belong to the adjoint 
representation of the gauge group, 
$\Phi$ and $F_\Phi$ can belong to any of the 
representations appearing in the symmetric product of the 
two adjoint  representations of corresponding gauge group. 
%_____________________________________________________________________________

Since the  SM can be embedded into a larger
gauge group
%, where an entire SM generation can be fitted into
%a single (ir)reducible representation of the underlying gauge group,
the question of unified gauge group needs to be discussed in order to
study the implications for nonuniversal soft gaugino masses in grand unified
theories. We recall that there is chain of group embeddings of the SM gauge 
group into a larger group~\cite{Ramond:1979py}
\be
{SU(3)_C \times SU(2)_L \times U(1)_Y \subset 
SU(5) \subset SO(10)\subset E_6\subset E_7 \subset E_8.}
\label{embed1}
\ee
However, we note that 
in four-dimensional grand unified theories the gauge groups
$E_7$ and $E_8$ do not lead to a chiral structure
of the weak interactions, and hence are ruled out 
as grand unified gauge groups on phenomenological grounds. 
This leaves out only
the three groups, $SU(5)$, $SO(10)$, and  $E_6$ as possible grand 
unified gauge groups in four dimensions. Here we shall study the 
implications of nonuniversal gaugino masses  in the case of
$SU(5)$ and $SO(10)$  grand unified gauge groups.
%_______________________________________________________________

\subsubsection{$SU(5)$}
\label{non-universal-SU(5)}
%_________________________________________________________________
In this Section we shall consider the case of embedding of the SM gauge group 
into the  grand unified gauge group $SU(5)$.
In the symmetric product of the two adjoint~({\bf 24} dimensional)
representations of $SU(5)$, we have
\bea
({\bf 24 \otimes 24})_{Symm} = {\bf 1 \oplus 24 \oplus 75 \oplus 200}.
\label{product}
\eea
In the simplest case where $\Phi$ and $F_\Phi$ are
assumed to be in the singlet representation of $SU(5)$,  we have
equal gaugino masses at the GUT scale.  But, as we can see  from
(\ref{product}), $\Phi$ and  $F_\Phi$
can belong to any of the non-singlet representations
{\bf 24}, {\bf 75},  and {\bf 200} of $SU(5)$. In such  cases 
the soft gaugino masses are unequal but related to one another via the
representation invariants of the gauge group~\cite{Ellis:1985jn}.  
In Table~\ref{tab1} we show the ratios of  gaugino masses 
which result when $F_{\Phi}$ belongs to different representations of 
$SU(5)$ in the decomposition~(\ref{product}) .
In this paper, for definiteness, we shall study the case of each 
representation independently, although an arbitrary combination of these is 
also allowed.

In the one-loop approximation, the solution of renormalization 
group~(RG) equations for the soft supersymmetry breaking
gaugino masses $M_1$, $M_2$, and $M_3$  can be written 
as~\cite{Martin:1993ft}
\bea {{M_i (t)} \over {\alpha_i(t)}} = {{{M_i}({\rm GUT})} \over
{{\alpha_i}({\rm GUT})}}, \, \, \, i = 1, 2, 3. 
\label{rel1} \eea
\noindent Then at  any arbitrary scale  we have :
\bea
{M_1} = {\frac 5 3}{{\alpha} \over {\cos^2{\theta_W}}}
\left({{{M_1}({\rm GUT})} \over {{\alpha_1}({\rm GUT})}}\right),\;\;
%\eea
%\bea
{M_2} = {{\alpha} \over {\sin^2{\theta_W}}}
\left({{{M_2}({\rm GUT})} \over {{\alpha_2}({\rm GUT})}}\right),\;\;
%\eea
%\bea
{M_3} = {\alpha_3} \left({{{M_3}({\rm GUT})} \over
{{\alpha_3}({\rm GUT})}}\right).\nonumber\\
\label{rel2}
\eea
\noindent With these results, we can write the gaugino masses 
for the  {\bf 24} dimensional representation of $SU(5)$ as 
\bea
{{M_1} \over {M_3}} = -\frac 12 \left({\frac 5 3}{{\alpha} \over
{\cos^2{\theta_W}}}\right) \left({1 \over {\alpha_3}}\right),\;\;
%\eea
%\bea
{{M_2} \over {M_3}} = -\frac 32\left({{\alpha} \over
{\sin^2{\theta_W}}}\right) \left({1 \over {\alpha_3}}\right).
\label{rel3}
\eea
\noindent Similarly, for the {\bf 75} dimensional representation of
$SU(5)$, we have the result
\bea
{{M_1} \over {M_3}} = -5 \left({\frac 5 3}{{\alpha} \over
{\cos^2{\theta_W}}}\right) \left({1 \over {\alpha_3}}\right),\;\;
%\eea
%\bea
{{M_2} \over {M_3}} = 3\left({{\alpha} \over
{\sin^2{\theta_W}}}\right) \left({1 \over {\alpha_3}}\right),
\label{rel4}
\eea
\noindent 
and finally for the {\bf 200} dimensional representation of $SU(5)$ we have
\bea
{{M_1} \over {M_3}} = 10 \left({\frac 5 3}{{\alpha} \over
{\cos^2{\theta_W}}}\right) \left({1 \over {\alpha_3}}\right),\;\;
{{M_2} \over {M_3}} = 2 \left({{\alpha} \over
{\sin^2{\theta_W}}}\right) \left({1 \over {\alpha_3}}\right).
\label{rel5}
  \eea 
%_______________________________________________________

\begin{table}[t!]
\renewcommand{\arraystretch}{1.0}
\begin{center}
%  \centering
  \begin{tabular}{||c|ccc|ccc||}
    \hline 
    $SU(5)$ & $M_1^G$ & $M_2^G$ & $M_3^G$ & 
    $M_1^{EW}$ & $M_2^{EW}$ & $M_3^{EW}$
    \\ \hline 
    {\bf 1} & 1 & 1
    & 1 & 1 & 2 & 7.1 \\ 
    & & & & & & \\    
    {\bf 24} & 1 & 3 & -2 & 1 & 6 & -14.3 \\
     & & & & & & \\    
     {\bf 75} & 1 &-$\frac{3}{5}$ &-$\frac{1}{5}$ & 1 & -1.18 & -1.41 \\
      & & & & & & \\    
      {\bf 200} & 1 & $\frac{1}{5}$ &$\frac{1}{10}$ &1 & 0.4 & 0.71
    \\ \hline
  \end{tabular}
  \end{center}
  \caption{\label{tab1} Ratios of the gaugino masses at the GUT scale
    in the normalization ${M_1}(GUT)$ = 1, and at the electroweak
    scale in the normalization ${M_1}(EW)$ = 1 
    for $F$-terms in different representations of $SU(5)$.
    These results are obtained by using 1-loop renormalization
    group equations.}
\renewcommand{\arraystretch}{1.0}
\end{table}
\noindent
To compute the ratios of the gaugino masses at the electroweak~(EW) 
scale $M_Z$ for different representations of $SU(5)$ 
in the product (\ref{product}), we use
the relevant renormalization group equations
for the soft gaugino masses.  In  Table~\ref{tab1}
we show the approximate results for these
masses at the electroweak scale $M_i(EW)$. 
These are calculated using one loop 
renormalization group equations for the gaugino masses and the gauge
couplings. The effect of two-loop calculations is to
increase the ratio $M_1/M_2$ by  an  amount which is not
significant. It is important to note  that
these results are consistent with the unification of gauge
couplings 
\bea \alpha^G_3 = \alpha^G_2 = \alpha^G_1 = \alpha^G
(\approx 1/25), \eea 
at the GUT scale, where we have
neglected the contribution of non universality in gaugino masses
to the gauge couplings, which is not significant.

\begin{figure}[ht]
\begin{minipage}[b]{0.45\linewidth}
\vspace*{0.65cm}
\centering
\includegraphics[width=6.5cm, height=5cm]{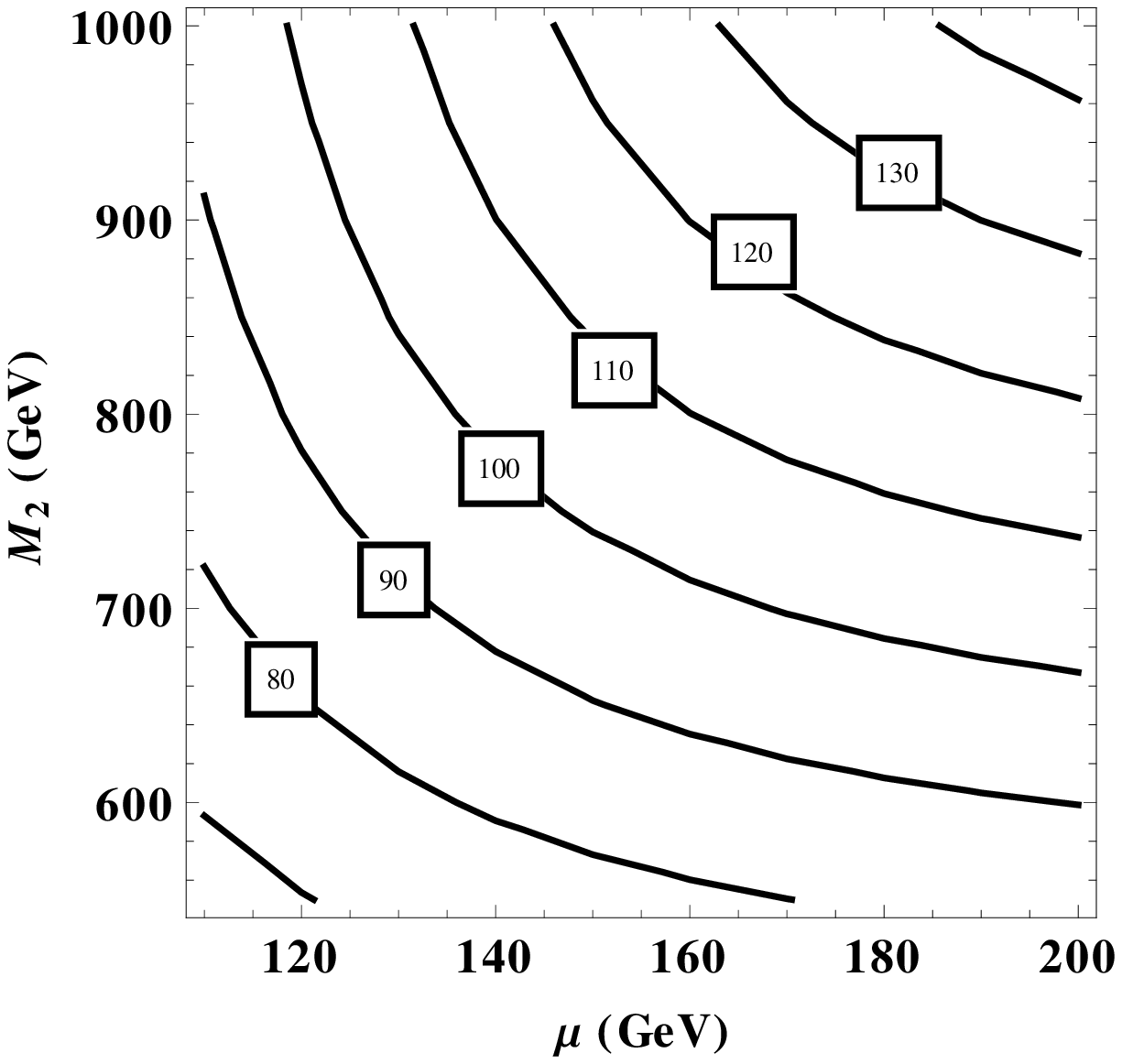}
\caption{The lightest neutralino mass in $\mu - M_2$ plane for $SU(5)$ with $\Phi$ and $F_\Phi$ in the
{\bf 24} dimensional representation. The value of $\mu$ 
giving $m_{\tilde {\chi}_1^0} \approx$ 108 GeV, for the smallest value of $M_2$ satisfying the 
gluino mass constraint, is chosen for our analyses.}
\label{fig:mum224}
\end{minipage}
\hspace{0.7cm}
\begin{minipage}[b]{0.45\linewidth}
\centering
\includegraphics[width=6.5cm, height=5cm]{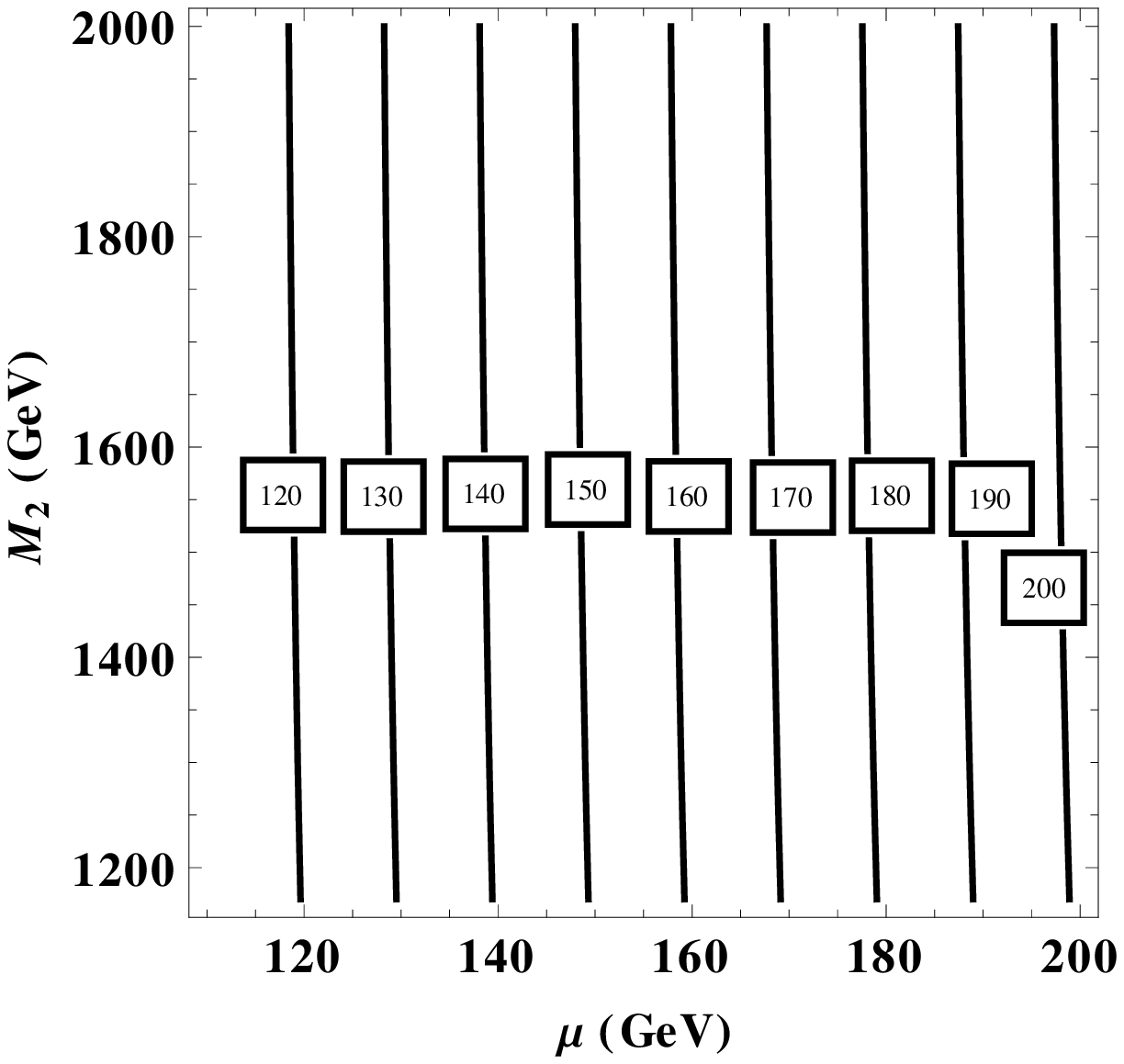}
\caption{The lightest neutralino mass in $\mu - M_2$ plane for $SU(5)$ with $\Phi$ and $F_\Phi$ in the
{\bf 75} dimensional representation. The value of $\mu$ 
giving $m_{\tilde {\chi}_1^0} \approx$ 108 GeV, for the smallest value of $M_2$ satisfying the gluino mass 
constraint, is chosen for our analyses. }
\label{fig:mum275}
\end{minipage}
\end{figure}

\begin{figure}[htb]
\centering
\includegraphics[width=6.5cm, height=5cm]{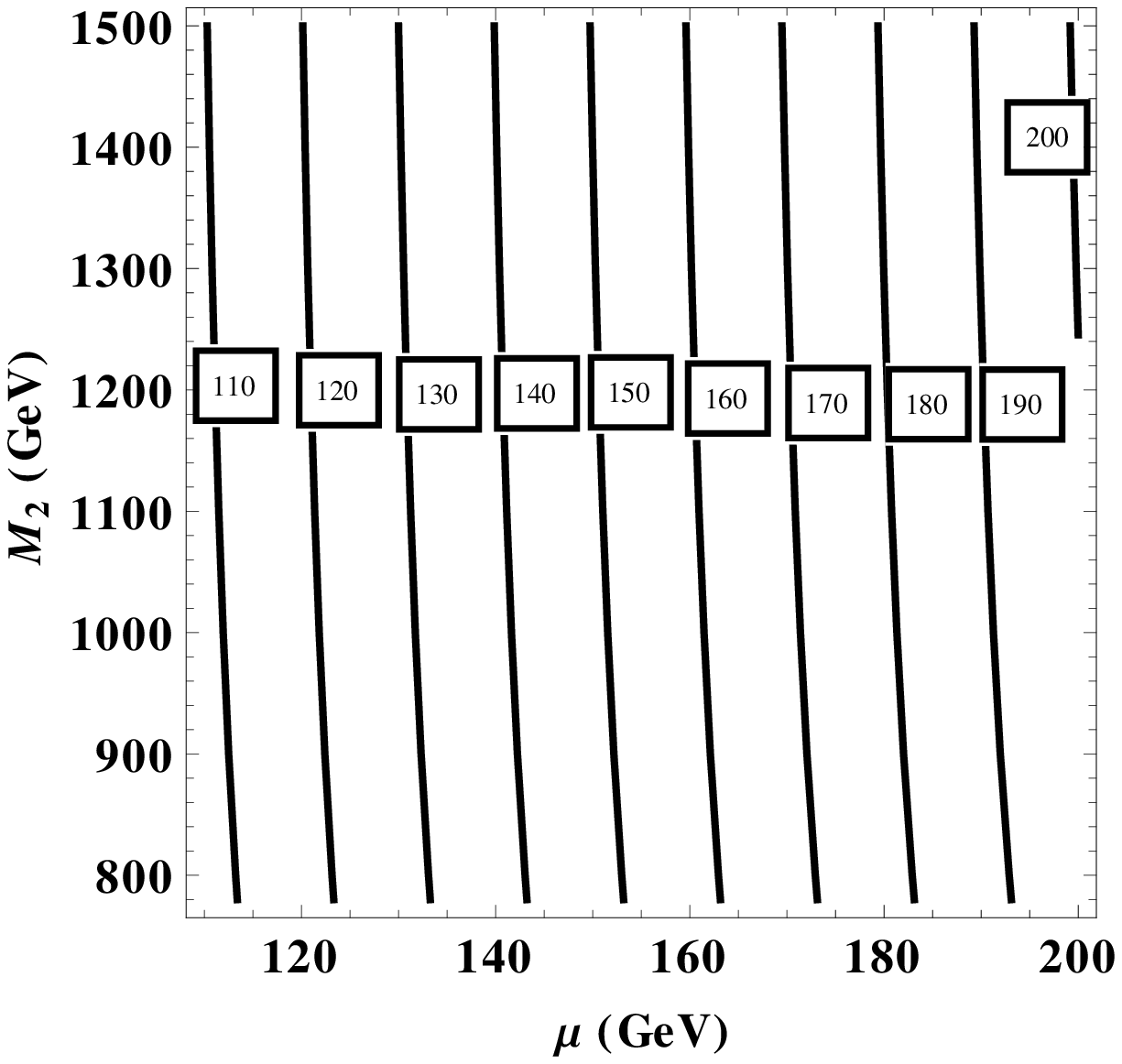}
\caption{The lightest neutralino mass in $\mu - M_2$ plane for $SU(5)$ with $\Phi$ and $F_\Phi$ in the
{\bf 200} dimensional representation. The value of $\mu$ 
giving $m_{\tilde {\chi}_1^0} \approx$ 108 GeV, for the smallest value of $M_2$ satisfying the 
gluino mass constraint, is chosen for our analyses.}
\label{fig:mum2200}
\end{figure}

%%%%% Table SU(5) Scenario for 24 representation %%%%%%%%
%
%\bigskip
\begin{table}[htb]
\renewcommand{\arraystretch}{1.0}
\begin{center}
\vspace{0.5cm}
        \begin{tabular}{|c|c|c|c|}
\hline
 $\tan\beta$ = 10 &$\mu$ = 138 GeV &$M_1$ = 149 GeV &$M_2$ = 890 GeV \\
\hline
 $M_3$ = -2121 GeV    &$A_t$= -1000 GeV &$A_b$= -2050 GeV &$A_{\tau}$= -2050 GeV \\
\hline
$m_{\chi_1^0}$ = 108 GeV &$m_{\chi_1^\pm}$ = 138.5 GeV
 &$m_{\tilde e_R}$ = 156 GeV &$m_{\tilde\nu_e}$ = 136  GeV \\
%\cline{1-3}
\hline
$m_{\chi_2^0}$ = 146 GeV &$m_{\chi_2^\pm}$ = 903 GeV
 &$m_{\tilde e_L}$ = 157 GeV &$m_h$ = 123.5 GeV \\
\hline
\end{tabular}
\end{center}
\vspace{-0.5cm}
\caption{Input parameters and resulting masses for various states in $SU(5)$
supersymmetric grand unified theory with $\Phi$ and $F_\Phi$ in the
{\bf 24} dimensional representation.} We shall refer to this  model
as $[SU(5)]_{24}$ in the text.
\renewcommand{\arraystretch}{1.0}
\label{parEWSB24}
\end{table}
%%%%% End table SU(5) Scenario for 24 representation  %%%%
%
%%%%% Table SU(5) Scenario for 75 representation %%%%%%%%
%
%\bigskip
\begin{table}[htb]
\renewcommand{\arraystretch}{1.0}
\begin{center}
\vspace{0.5cm}
        \begin{tabular}{|c|c|c|c|}
\hline
 $\tan\beta$ = 10 &$\mu$ = 108 GeV &$M_1$ = 993 GeV &$M_2$ = -1172 GeV \\
\hline
 $M_3$ = -1400 GeV    &$A_t$= 1000 GeV &$A_b$= 2300 GeV &$A_{\tau}$= 2300 GeV \\
\hline 
$m_{\chi_1^0}$ = 108 GeV &$m_{\chi_1^\pm}$ = 109 GeV 
 &$m_{\tilde e_R}$ = 156 GeV &$m_{\tilde\nu_e}$ = 136  GeV \\
%\cline{1-3}
\hline
$m_{\chi_2^0}$ = 110 GeV &$m_{\chi_2^\pm}$ = 1916 GeV  
 &$m_{\tilde e_L}$ = 157 GeV &$m_h$ = 123.5 GeV \\ 
\hline
\end{tabular}
\end{center}
\vspace{-0.5cm}
\caption{Input parameters and resulting masses for various states in $SU(5)$
supersymmetric grand unified theory with $\Phi$ and $F_\Phi$ in the
{\bf 75} dimensional representation.}  We shall refer to this  model
as $[SU(5)]_{75}$ in the text.

\renewcommand{\arraystretch}{1.0}
\label{parEWSB75}
\end{table}
%%%%% End table SU(5) Scenario for 75 representation  %%%%
%
%
%%%%% Table SU(5) Scenario for 200 representation %%%%%%%%
%
%\bigskip
\begin{table}[htb]
\renewcommand{\arraystretch}{1.0}
\begin{center}
\vspace{0.5cm}
        \begin{tabular}{|c|c|c|c|}
\hline
 $\tan\beta$ = 10 &$\mu$ = 111 GeV &$M_1$ = 1970 GeV &$M_2$ = 788 GeV \\
\hline
 $M_3$ = 1399 GeV    &$A_t$= 1000 GeV &$A_b$= 2300 GeV &$A_{\tau}$= 2300 GeV \\
\hline 
$m_{\chi_1^0}$ = 107.7 GeV &$m_{\chi_1^\pm}$ = 111 GeV 
 &$m_{\tilde e_R}$ = 166 GeV &$m_{\tilde\nu_e}$ = 136 GeV \\
%\cline{1-3}
\hline
$m_{\chi_2^0}$ = 117 GeV &$m_{\chi_2^\pm}$ = 807 GeV  
 &$m_{\tilde e_L}$ = 157 GeV &$m_h$ = 123.7 GeV \\ 
\hline
\end{tabular}
\end{center}
\vspace{-0.5cm}
\caption{Input parameters and resulting masses for various states in $SU(5)$
supersymmetric grand unified theory with $\Phi$ and $F_\Phi$ in the
{\bf 200} dimensional representation.}  We shall refer to this  model
as $[SU(5)]_{200}$ in the text.
\renewcommand{\arraystretch}{1.0}
\label{parEWSB200}
\end{table}
%%%%%%%%%%%%%%%%%% End table SU(5) Scenario for 200 representation  %%%%
The input parameters for $\bf 24$, $\bf 75$ and $\bf 200$ 
dimensional representations of $SU(5)$
in Table \ref{tab1} are chosen in a manner similar 
to the one we used for the parameter space of MSSM EWSB.
The lowest values of $M_1$ and $M_2$ satisfying the gluino mass
constraint are chosen from Fig. \ref{fig:m1m2} corresponding
to various $SU(5)$ representations shown there. We then 
plot the values of $M_2$ versus $\mu$ for various
representations in  contour plots shown in Figs.~\ref{fig:mum224}, \ref{fig:mum275} and \ref{fig:mum2200}.
The values of $M_2$ and $\mu$  are then selected which result in 
the lightest neutralino mass of $108$ GeV and chargino 
mass larger than $110$ GeV. For $SU(5)_{75}$ and  $SU(5)_{200}$ the bino
and wino mass parameters $M_1, M_2 \gg \mu$, therefore the LSP mass in these cases is almost equal to $\mu$. This is seen
from Figs.~\ref{fig:mum275} and \ref{fig:mum2200}, where the contours in the $M_2 - \mu$ plane are independent of $M_2$.

The input parameters and the resulting masses for the $\bf 24$,
$\bf 75$, and $\bf 200$ dimensional representations of $SU(5)$ 
which result in nonuniversal gaugino masses at the grand unified 
scale obtained in a manner described above are shown in 
Tables~\ref{parEWSB24}, \ref{parEWSB75} 
and \ref{parEWSB200}, respectively. 
In  arriving at the parameter values in these Tables, 
we have
taken into account various theoretical and phenomenological constraints,
including the electroweak symmetry breaking at the correct scale,
as described  in the Appendix \ref{exp_cons}. Other values can
be obtained by choosing larger values of the parameter $M_3$.  

The composition of the lightest neutralino for the different 
representations of $SU(5)$ in Table~\ref{tab1} is obtained from the
mixing matrix for the choices of parameters given in Tables~
\ref{parEWSB24}, \ref{parEWSB75} and \ref{parEWSB200}. 
This composition is calculated to be:
\begin{enumerate}
\item
$SU(5)$ with $\Phi$ and $F_\Phi$ in the
{\bf 24} dimensional representation~(labelled as model $[SU(5)]_{24}$):
\begin{eqnarray}
N_{1j} & = & (0.638,~ -0.054,~ 0.598,~ -0.482);
\label{SU(5)24}
\end{eqnarray}
\item
$SU(5)$ with $\Phi$ and $F_\Phi$ in the
{\bf 75} dimensional representation~(labelled as model  $[SU(5)]_{75}$):
\begin{eqnarray}
N_{1j} & = & (0.031,~ 0.056,~ -0.710,~ -0.700);
\label{SU(5)75}
\end{eqnarray}
\item
$SU(5)$ with $\Phi$ and $F_\Phi$ in the
{\bf 200} dimensional representation~(labelled as model  $[SU(5)]_{200}$):
\begin{eqnarray}
N_{1j} & = & (0.018,~ -0.085,~ 0.719,~ -0.689).
\label{SU(5)200}
\end{eqnarray}
\end{enumerate}

We note from (\ref{SU(5)24}),  (\ref{SU(5)75}),  and (\ref{SU(5)200})
that for the $\bf 24$ dimensional representation of $SU(5)$, the 
dominant component of the neutralino is the bino, 
whereas for the other representations of $SU(5)$, 
there is a higgsino like lightest neutralino. Thus, 
for $\bf 75$ and $\bf 200$ dimensional representations
the neutralino being higgsino like couples weakly to the selectron, 
with the dominant contribution to the cross section coming from the 
neutralino- $Z^0$ coupling. 

%______________________________________________
\subsubsection{$SO(10)$}
\label{non-universal-SO(10)}
We now consider the embedding of the SM gauge group in a
$SO(10)$  supersymmetric grand unified theory.
Since the adjoint representation of $SO(10)$ is {\bf 45} dimensional, 
$\Phi$ and $F_\Phi$ can belong to any of the following representations
appearing~\cite{Martin:2009ad} in the symmetric product of two $\bf 45$ 
dimensional representations of $SO(10)$:
%_______
\bea 
({\bf 45} \times {\bf 45})_{Symm}={\bf 1} 
\oplus {\bf 54} \oplus {\bf 210} \oplus {\bf 770}.
\label{symmetric_SO10}
\eea
%_______
There are three maximal proper subgroups of $SO(10)$ which are consistent 
with the fermion content of the Standard Model. 
These are ($i$) $SU(5) \subset SO(10)$  
with  the normal~(nonflipped) embedding; 
($ii$) $SU(5)' \times U(1) \subset SO(10)$  with the flipped
embedding; and ($iii$) $SU(4) \times SU(2)_L \times SU(2)_R \subset SO(10)$ 
embedding.
Using  relations (\ref{rel1}) and (\ref{rel2}),  we obtain the gaugino mass 
parameters
at the GUT scale for different representations that arise in the symmetric
product of two adjoint representations of $SO(10)$ with  the relevant 
embedding of the SM gauge group in $SO(10)$.
These are shown in 
Tables~ \ref{tab2}, \ref{tab3}, and \ref{tab4}.
The ratio of the gaugino masses 
at the GUT scale for the different cases  for $SO(10)$  shown in these
Tables can be scaled 
down to the electroweak scale, as described above.
The results for the gaugino masses at the electroweak scale for these cases are 
also shown in Tables~\ref{tab2}, \ref{tab3}, and  \ref{tab4}.

%________________________________________________________________
\begin{table}[ht]
\begin{minipage}[b]{0.45\linewidth}
  \begin{tabular}{||c|c|ccc|ccc||}
   \hline 
   $SO(10)$ & $SU(5)$ & $M_1^G$ & $M_2^G$ & $M_3^G$ & 
    $M_1^{EW}$ & $M_2^{EW}$ & $M_3^{EW}$
    \\ \hline 
    {\bf 1} & {\bf 1} &1 &1 &1 &1 &2 &7.1\\[0.5 mm]
%     & & & & & & & \\
     {\bf 54} & {\bf 24} &1 &3 &-2 
     & 1 &6  &-14.3 \\[0.5 mm]
%       & & & & & & & \\
 {\bf 210} & {\bf 1}  &1 &1 &1 &1 &2 &7.1\\[0.5 mm]
%   & & & & & & & \\
           &  {\bf 24} &1 &3 &-2
     & 1 & 6  &-14.3 \\[0.5 mm]
%      & & & & & & & \\ 
           & {\bf 75} & 1 &-$\frac{3}{5}$ &-$\frac{1}{5}$ & 1 & -1.18 & -1.41 \\[0.5 mm]
%        & & & & & & & \\    
  {\bf 770} & {\bf 1}  &1 &1 &1 &1 &2 &7.1\\[0.5 mm]
%     & & & & & & & \\    
             &  {\bf 24} &1 &3 &-2
     & 1 &6  &-14.3 \\[0.5 mm]
 %       & & & & & & & \\    
           & {\bf 75} & 1 &-$\frac{3}{5}$ &-$\frac{1}{5}$ & 1 & -1.18 & -1.14 \\[0.5 mm]
%            & & & & & & & \\      
           & {\bf 200} & 1 & $\frac{1}{5}$ &$\frac{1}{10}$ &1 & 0.4 & 0.71          
         \\ \hline
  \end{tabular}
   \caption{\label{tab2}Ratios of the gaugino masses at the GUT scale
    in the normalization ${M_1}(GUT)$ = 1, and at the electroweak
    scale in the normalization ${M_1}(EW)$ = 1 
    for $F$-terms in representations of $SU(5)\subset SO(10)$
    with the normal (nonflipped) embedding. These results have been 
    obtained at the 1-loop level.}
 \end{minipage}\qquad
\hspace{0.2cm}
\begin{minipage}[b]{0.49\linewidth}
\begin{center}
  \begin{tabular}{||c|c|ccc|ccc||}
   \hline 
   $SO(10)$ & $[SU(5)' \times U(1)]_{flipped}$ & $M_1^G$ & $M_2^G$ & $M_3^G$ & 
    $M_1^{EW}$ & $M_2^{EW}$ & $M_3^{EW}$
    \\ \hline 
    {\bf 1} & ({\bf 1},0) &1 &1 &1 &1 &2 &7.1\\ [0.5 mm]
 %     & & & & & & & \\
     {\bf 54} & ({\bf 24},0) & 1 &3 &-2 
     & 1 &6  &-14.3 \\ [0.5 mm]
%      & & & & & & & \\
    {\bf 210} & ({\bf 1},0)  & 1 &-$\frac{5}{19}$ &-$\frac{5}{19}$ &1 &-0.52 &-1.85\\ [0.5 mm]
 %    & & & & & & & \\
           &  ({\bf 24},0) & 1 &-$\frac{15}{7}$ &$\frac{10}{7}$ 
     & 1 &-4.2  &10 \\ [0.5 mm]
%      & & & & & & & \\
           & ({\bf 75},0) & 1 &-15 & -5 & 1 &-28 & -33.33 \\ [0.5 mm]
%       & & & & & & & \\     
  {\bf 770} & ({\bf 1},0)  & 1 &$\frac{5}{77}$  &$\frac{5}{77}$ &1 &0.13 &0.46\\ [0.5 mm]
%   & & & & & & & \\
             &  ({\bf 24},0) & 1 &$\frac{15}{101}$ & -$\frac{10}{101}$  
     & 1 &0.3  &-0.70 \\ [0.5 mm]
%     & & & & & & & \\ 
           & ({\bf 75},0) & 1 &-15 &-5 & 1 &-28 &-33.3 \\ [0.5 mm]
%     & & & & & & & \\       
           & ({\bf 200},0) & 1 &5 & $\frac{5}{2}$ &1 & 9.33 & 16.67          
         \\ \hline
  \end{tabular}
  \end{center}
  \caption{\label{tab3}Ratios of the gaugino masses at the GUT scale
    in the normalization ${M_1}(GUT)$ = 1, and at the electroweak
    scale in the normalization ${M_1}(EW)$ = 1 at the 1-loop level
    for $F$-terms in representations of flipped
    $SU(5)'\times U(1)$ $\subset SO(10)$.}
\end{minipage}

\vspace{1.0cm}

\begin{minipage}[b]{0.55\linewidth}
\begin{center}
   \begin{tabular}{||c|c|ccc|ccc||}
   \hline 
   $SO(10)$ & $ SU(4) \times SU(2)_R $ & 
                                          $M_1^G$ & $M_2^G$ & $M_3^G$ & 
    $M_1^{EW}$ & $M_2^{EW}$ & $M_3^{EW}$
    \\ \hline 
    {\bf 1} & ({\bf 1},{\bf 1}) &1 &1 &1 &1 &2 &7.1\\ [0.5 mm]
%     & & & & & & & \\
     {\bf 54} & ({\bf 1},{\bf 1}) & 1 &3 &2  
     & 1 &6  &-14.3 \\ [0.5 mm]
 %    & & & & & & & \\
    {\bf 210} & ({\bf 1},{\bf 1})  & 1 &-$\frac{5}{3}$ &0 &1 &-3.35 &0\\ [0.5 mm]
%    & & & & & & & \\
           &  ({\bf 15},{\bf 1}) & 1 &0 &-$\frac{5}{4}$
     & 1 &0  &-9.09 \\ [0.5 mm]
 %    & & & & & & & \\
           & ({\bf 15},{\bf 3}) & 1 & 0 & 0 &1  & 0 &0 \\ [0.5 mm]
%      & & & & & & & \\     
  {\bf 770} & ({\bf 1},{\bf 1})  & 1 &$\frac{25}{19}$ &$\frac{10}{19}$&1 &2.6 &3.7\\ [0.5 mm]
%  & & & & & & & \\
             &  ({\bf 1},{\bf 5}) & 1 &0 &0  
     & 1 &0  &0 \\ [0.5 mm]
 %   & & & & & & & \\ 
           & ({\bf 15},{\bf 3}) & 1 & 0 & 0 & 1 & 0 & 0 \\ [0.5 mm]
%    & & & & & & & \\       
           & ({\bf 84},{\bf 1}) & 1 & 0 & $\frac{5}{32}$ &1 & 0 & 1.11          
         \\ \hline 
  \end{tabular}
  \end{center}
   \caption{\label{tab4}Ratios of the gaugino masses at the GUT scale
    in the normalization ${M_1}(GUT)$ = 1, and at the electroweak
    scale in the normalization ${M_1}(EW)$ = 1 at the 1-loop level
    for $F$-terms in representations of 
    $SU(4)\times SU(2)_L \times SU(2)_R \subset SO(10)$.} 
  \end{minipage}
  \end{table}

We note from Table~\ref{tab2} that the ratios of gaugino masses
for the different representations of $SO(10)$ in the symmetric product
(\ref{symmetric_SO10}) with the unflipped embedding 
$SU(5) \subset SO(10)$ are identical to
the corresponding gaugino mass ratios in Table~\ref{tab1} for 
the embedding of SM in $SU(5)$. Therefore, the input parameters and the 
resulting masses for the gaugino mass ratios in Table~\ref{tab2} for
$SO(10)$ are identical
to the corresponding  Tables~\ref{parEWSB24}, \ref{parEWSB75}, 
and \ref{parEWSB200} for $SU(5)$. 

On the other hand, for the flipped embedding 
$SU(5)' \times U(1) \subset SO(10)$, Table~\ref{tab3},
the gaugino mass ratios for the $\bf 210$ and $\bf 770$ dimensional 
representations of the grand unified gauge group
can be different from the corresponding ratios
for $SU(5)$.

\begin{figure}[ht]
\includegraphics[width=6.5cm, height=5cm]{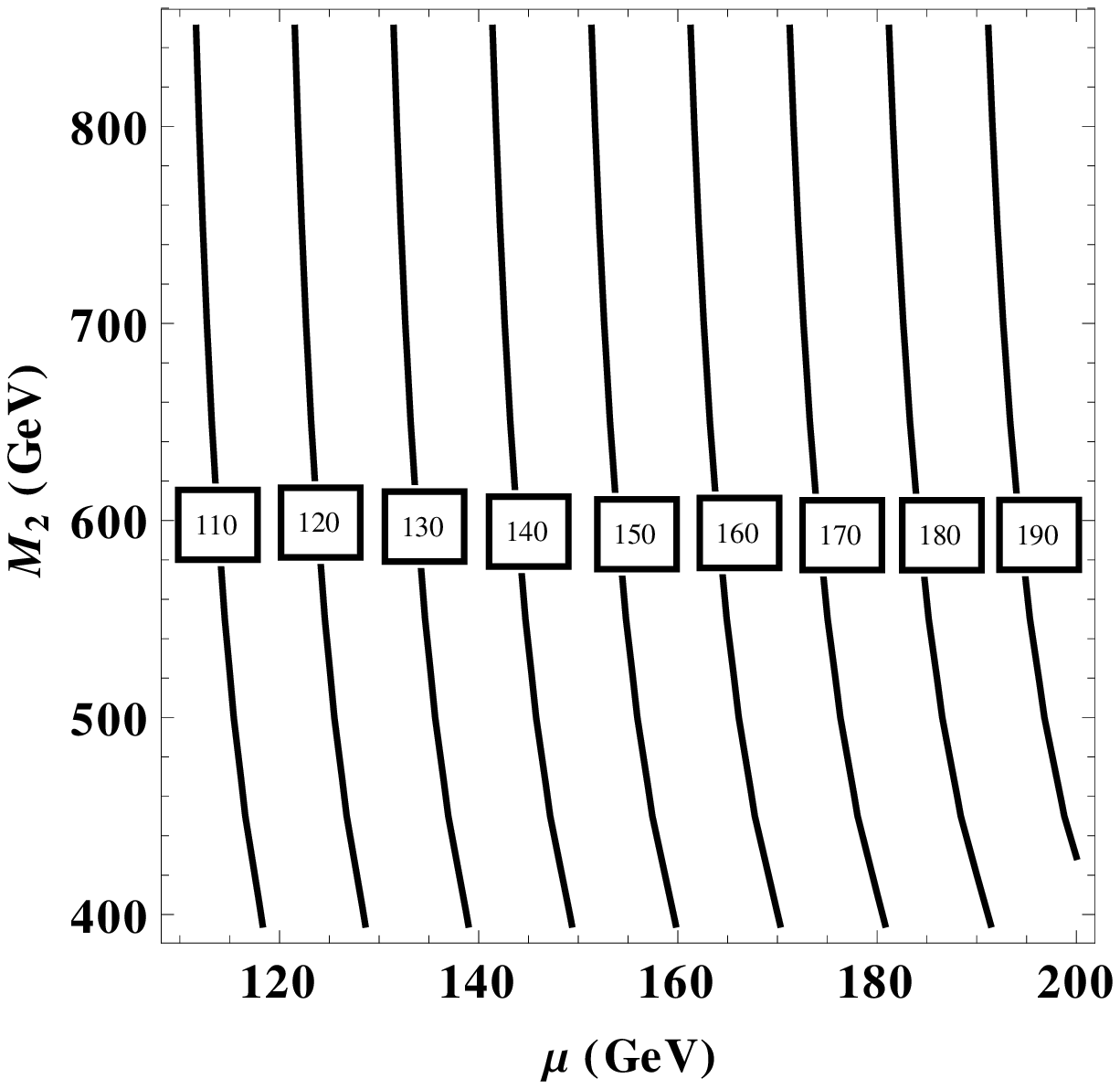}
\caption{The lightest neutralino mass in $\mu - M_2$ plane for $SO(10)$ where $SU(5)' \times U(1) \subset SO(10)$ 
with $\Phi$ and $F_\Phi$ in the {\bf 210} (~\ref{SO(10))210_1}). The value of $\mu$ 
giving $m_{\tilde {\chi}_1^0} \approx$ 108 GeV, for the selected value of $M_2$ is considered for our analyses. }
\label{fig:mum2210}
\end{figure}

\begin{figure}[ht]
\begin{minipage}[b]{0.45\linewidth}
\vspace*{0.65cm}
\centering
\includegraphics[width=6.5cm, height=5cm]{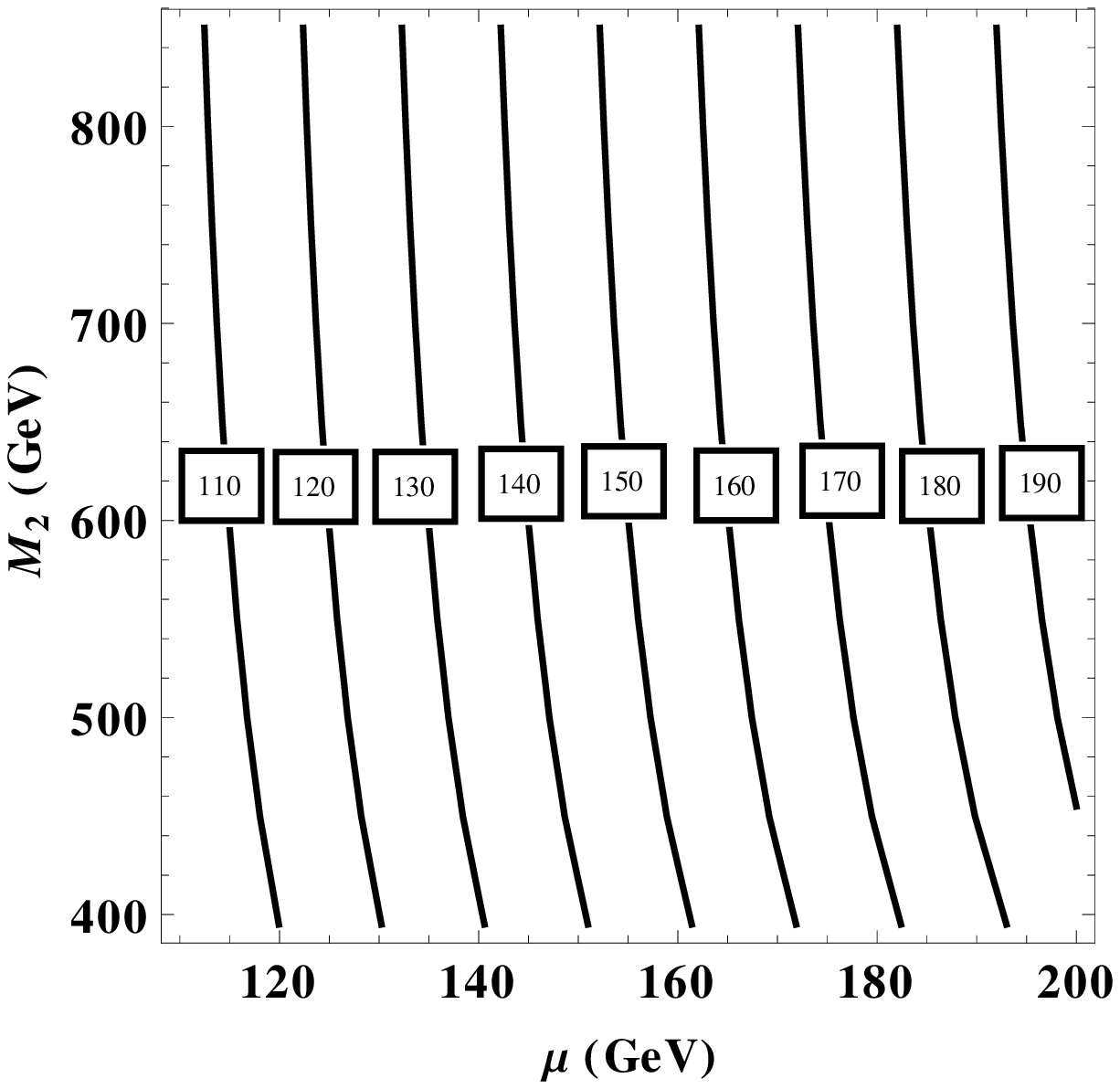}
\caption{The lightest neutralino mass in $\mu - M_2$ plane for $SO(10)_{770}$ scenario. The value of $\mu$ 
giving $m_{\tilde {\chi}_1^0} \approx$ 108 GeV, for the smallest value of $M_2$ satisfying the 
gluino mass constraint, is chosen for our analyses.}
\label{fig:mum27701}
\end{minipage}
\hspace{0.7cm}
\begin{minipage}[b]{0.45\linewidth}
\centering
\includegraphics[width=6.5cm, height=5cm]{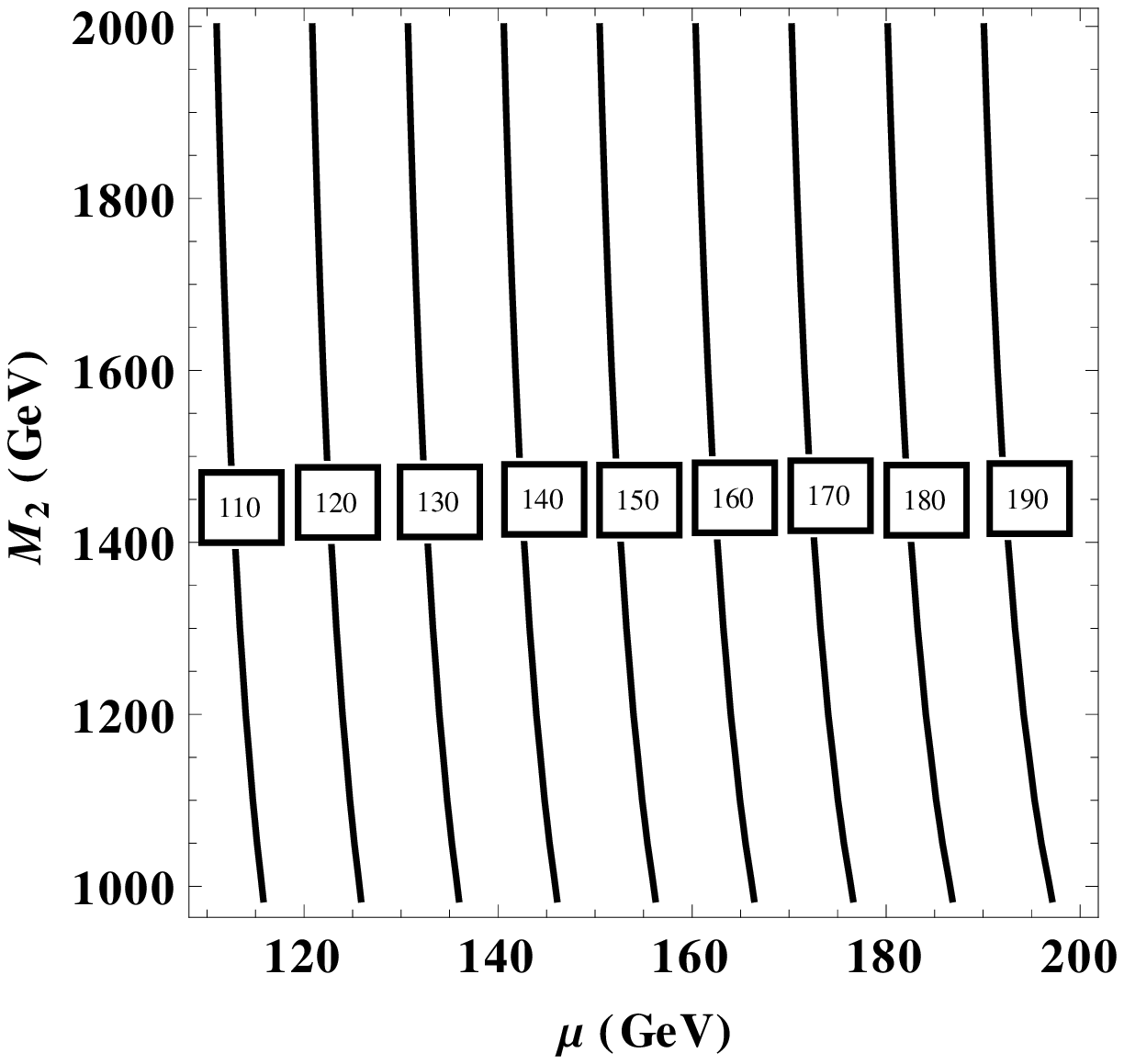}
\caption{The lightest neutralino mass in $\mu - M_2$ plane for $SO(10)_{770'}$ scenario. The value of $\mu$ 
giving $m_{\tilde {\chi}_1^0} \approx$ 108 GeV, for the smallest value of $M_2$ satisfying the gluino mass 
constraint, is chosen for our analyses. }
\label{fig:mum27702}
\end{minipage}
\end{figure}

In the case of $SO(10$ we choose the parameter values in a manner
similar to the case of MSSM EWSB and the $SU(5)$ grand unified theory.
The result of such a procedure is shown in 
Fig. \ref{fig:m1m2}, and in Figs.\ref{fig:mum2210}, \ref{fig:mum27701} and \ref{fig:mum27702}.
The resulting values of parameters are shown in Tables~\ref{parSO(10)_210_1}, \ref{parSO(10)_770_1}
and \ref{parSO(10)_770_2}.

We  note that the ratio of the gaugino masses  for the $\bf (24, 0),
(75, 0)$ of the $\bf 210$ dimensional representation of
$SO(10)$, and $\bf (75, 0), (200, 0)$ of the
$\bf 770$ representation of $SO(10)$ follow the same
pattern as the $\bf (24, 0)$ of the $\bf 54$
dimensional representation of $SO(10)$. 
The behaviour of the cross section for the radiative neutralino cross
section will be similar in these cases due to the fact that the 
dominant component
of the lightest neutralino in this case is a bino.
Similarly, the $\bf (24, 0)$ of the $\bf 770$ dimensional
representation of $SO(10)$ has the same pattern as
$\bf 200$ dimensional representation of $SU(5)$.
Here also the behavior will be same. Because of this we shall 
focus on the $\bf (1, 0)$ option for the
$\bf 210$ and $\bf 770$ dimensional representations in Table~\ref{tab3}.
Similarly, because of phenomenological
reasons,  we focus on the
$\bf (1, 1)$  of  $SU(4) \times SU(2)_R$ of the $\bf 770$ 
representation only.

%

%%%%% Table S0(10) flipped Scenario for 210 representation %%%%%%%%
%
\begin{table}[htb]
\renewcommand{\arraystretch}{1.0}
\begin{center}
\vspace{0.5cm}
        \begin{tabular}{|c|c|c|c|}
\hline
 $\tan\beta$ = 10 &$\mu$ = 116 GeV &$M_1$ = 760 GeV &$M_2$ = -395 GeV \\
\hline
 $M_3$ = -1405 GeV    &$A_t$= 1000 GeV &$A_b$= 2300 GeV &$A_{\tau}$= 2300 GeV \\
\hline 
$m_{\chi_1^0}$ = 108 GeV &$m_{\chi_1^\pm}$ = 111 GeV 
 &$m_{\tilde e_R}$ = 156 GeV &$m_{\tilde\nu_e}$ = 136 GeV \\
%\cline{1-3}
\hline
$m_{\chi_2^0}$ = 122 GeV &$m_{\chi_2^\pm}$ = 422 GeV  
 &$m_{\tilde e_L}$ = 158 GeV &$m_h$ = 124.2 GeV \\ 
\hline
\end{tabular}
\end{center}
\vspace{-0.5cm}
\caption{Input parameters and resulting masses for various states in $SU(5)' \times U(1) \subset SO(10)$
supersymmetric grand unified theory with $\Phi$ and $F_\Phi$ in the
{\bf 210} dimensional representation with 
$SU(5)' \times U(1)$ in ({\bf 1},{\bf 0}) dimensional representation.}
We shall refer to this model as $[SO(10)]_{210}$ in the text.
\renewcommand{\arraystretch}{1.0}
\label{parSO(10)_210_1}
\end{table}
%%%%%%%%% End table SU(5)\subset SO(10) Scenario for 210 representation  %%%%
%%%%%%%%%%%%%%%%%%%%%%%%%%%
%
%%%%% Table S0(10) flipped Scenario for 770 representation %%%%%%%%
%
\begin{table}[htb]
\renewcommand{\arraystretch}{1.0}
\begin{center}
\vspace{0.5cm}
        \begin{tabular}{|c|c|c|c|}
\hline
 $\tan\beta$ = 10 &$\mu$ = 118 GeV &$M_1$ = 3038 GeV &$M_2$ = 395 GeV \\
\hline
 $M_3$ = 1398 GeV    &$A_t$= 1000 GeV &$A_b$= 2300 GeV &$A_{\tau}$= 2300 GeV \\
\hline 
$m_{\chi_1^0}$ = 108 GeV &$m_{\chi_1^\pm}$ = 113 GeV 
 &$m_{\tilde e_R}$ = 156 GeV &$m_{\tilde\nu_e}$ = 136 GeV \\
%\cline{1-3}
\hline
$m_{\chi_2^0}$ = 125 GeV &$m_{\chi_2^\pm}$ = 422 GeV  
 &$m_{\tilde e_L}$ = 157 GeV &$m_h$ = 124 GeV \\ 
\hline
\end{tabular}
\end{center}
\vspace{-0.5cm}
\caption{Input parameters and resulting masses for various states in $SU(5)' \times U(1) \subset SO(10)$
supersymmetric grand unified theory with $\Phi$ and $F_\Phi$ in the
{\bf 770} dimensional representation with $SU(5)' \times U(1)$ in ({\bf 1},{\bf 0})
dimensional representation.} We shall refer to this model as $[SO(10)]_{770}$
in the text.
\renewcommand{\arraystretch}{1.0}
\label{parSO(10)_770_1}
\end{table}
%%%%%%%%%%%%%%%%%% End table SO(10) flipped Scenario for 770 representation  %%%%
%%%%% Table S0(10) Scenario for 770 representation %%%%%%%%
%
\begin{table}[htb]
\renewcommand{\arraystretch}{1.0}
\begin{center}
\vspace{0.5cm}
        \begin{tabular}{|c|c|c|c|}
\hline
 $\tan\beta$ = 10 &$\mu$ = 113 GeV &$M_1$ = 378 GeV &$M_2$ = 985 GeV \\
\hline
 $M_3$ = 1402 GeV    &$A_t$= 1000 GeV &$A_b$= 2300 GeV &$A_{\tau}$= 2300 GeV \\
\hline 
$m_{\chi_1^0}$ = 107.4 GeV &$m_{\chi_1^\pm}$ = 114 GeV 
 &$m_{\tilde e_R}$ = 156 GeV &$m_{\tilde\nu_e}$ = 136 GeV \\
%\cline{1-3}
\hline
$m_{\chi_2^0}$ = 120 GeV &$m_{\chi_2^\pm}$ = 1000 GeV  
 &$m_{\tilde e_L}$ = 157 GeV &$m_h$ = 123.5 GeV \\ 
\hline
\end{tabular}
\end{center}
\vspace{-0.5cm}
\caption{Input parameters and resulting masses for various states in 
$SU(4) \times SU(2)_R \times SU(2)_L \subset SO(10)$
supersymmetric grand unified theory with $\Phi$ and $F_\Phi$ in the
{\bf 770} dimensional representation with $SU(4) \times SU(2)_R$ in ({\bf 1},{\bf 1})
dimensional representation.} We shall refer to this  model 
as $[SO(10)]_{770'}$ in the text.
\renewcommand{\arraystretch}{1.0}
\label{parSO(10)_770_2}
\end{table}
%%%%%%%%%%%%%%%%%% End table SO(10) Scenario for 770 representation  %%%%

 For the parameters of Tables~\ref{parSO(10)_210_1},  \ref{parSO(10)_770_1} and
\ref{parSO(10)_770_2}  the composition of the lightest neutralino is given
by the following.
%______
\begin{enumerate}
\item 
$SO(10)$ where $SU(5)' \times U(1) \subset SO(10)$ with $\Phi$ and $F_\Phi$ 
in the {\bf 210} dimensional representation with 
$SU(5)' \times U(1)$ in ({\bf 1},{\bf 0}) dimensional 
representation~(labelled as model $[SO(10)]_{210}$): 
\begin{eqnarray}
N_{1j} & = & (0.038,~ 0.194,~ -0.719,~ -0.666);  
\label{SO(10))210_1}
\end{eqnarray}
\item 
$SO(10)$ where $SU(5)' \times U(1) \subset SO(10)$ with 
$\Phi$ and $F_\Phi$ in the {\bf 770} dimensional representation 
with $SU(5)' \times U(1)$ in ({\bf 1},{\bf 0}) 
dimensional representation~(labelled as model $[SO(10)]_{770}$): 
\begin{eqnarray}
N_{1j} & = & (0.011,~ -0.193,~ 0.724,~- 0.663);  
\label{SO(10))770_1}
\end{eqnarray}
\item 
$SO(10)$ where $SU(4) \times SU(2)_R \times SU(2)_L \subset SO(10)$ with $\Phi$ and $F_\Phi$ in the {\bf 770} dimensional representation 
with $SU(4) \times SU(2)_R$ in ({\bf 1},{\bf 1})
dimensional representation~(labelled as model $[SO(10)]_{770'}$): 
\begin{eqnarray}
N_{1j} & = & (0.125,~ -0.660,~ 0.721,~- 0.678),  
\label{SO(10))770_2}
\end{eqnarray}
\end{enumerate}
implying thereby that higgsino  is the dominant component for the $\bf 210$ 
and $\bf 770$ dimensional representations with the embedding
$SU(5)' \times U(1) \subset SO(10)$ and for the $\bf 770$ dimensional representation with the
embedding $SU(4) \times SU(2)_R \times SU(2)_L \subset SO(10)$.
From the  Figs.\ref{fig:mum2210}, \ref{fig:mum27701} and \ref{fig:mum27702}
we see that, because the lightest neutralino is dominantly a higgsino,
the contours are almost independent of $M_2$.  

Thus, in these cases the dominant contribution to the radiative
neutralino production cross section will come from the 
neutralino-$Z^0$ coupling. Since the LSP for most of the scenarios considered here has a dominant higgsino component, 
the $Z^0$ width imposes a strict constraint, as
the $Z^0$ decay rate involves coupling to the higgsino component of the neutralino.
We have imposed the LEP constraint on the anomalous $Z^0$ decay width in our calculations :
\begin{equation}
 \Gamma(Z \rightarrow \tilde \chi_1^0 \tilde \chi_1^0) < 3 {\rm MeV}.
\end{equation}

%_________________________________________________________________________
\section{Radiative Neutralino Production in Grand Unified Theories}
\label{sec:radiative cross section}
\begin{figure}[h!]
{\unitlength=1.0 pt
\SetScale{1.0}
\SetWidth{1.0}      
\includegraphics{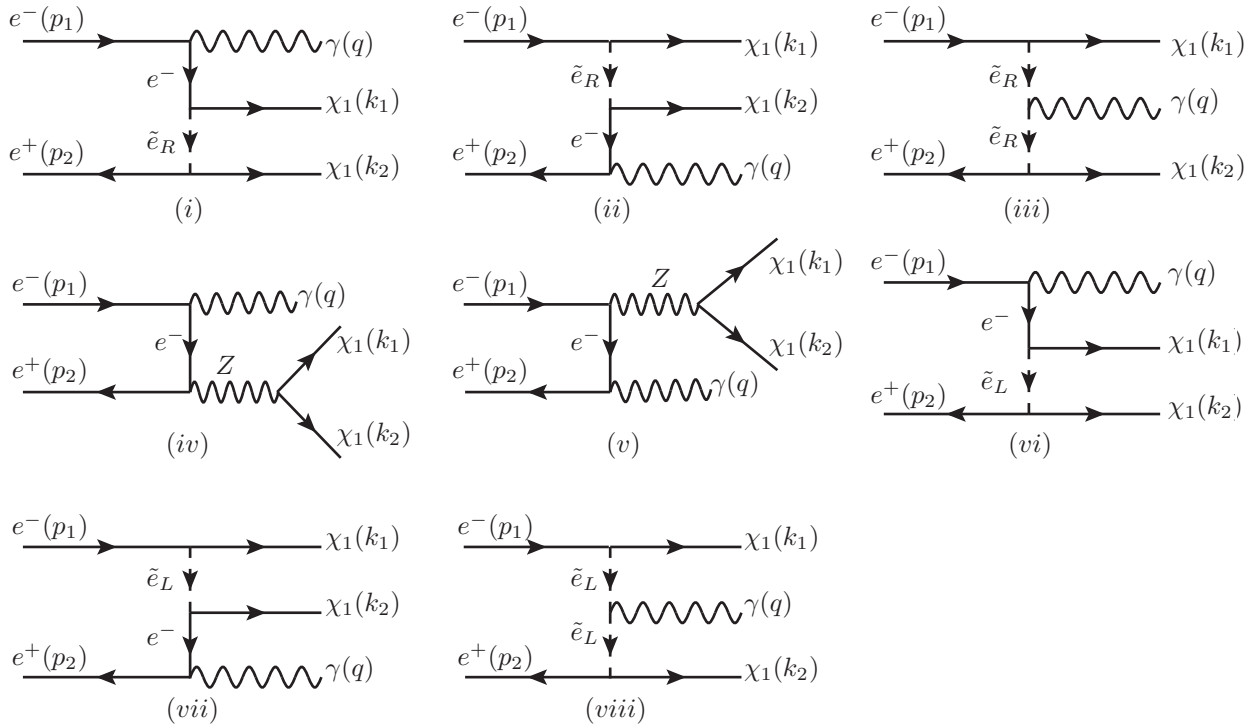}
}
\caption{Feynman diagrams contributing to the  radiative neutralino production 
 $e^+e^- \to  \tilde\chi_1^0\tilde\chi_1^0\gamma.$ There are six other diagrams
which are exchange diagrams corresponding to 
($i$, $ii$, $iii$, $vi$, $vii$, $viii$), with
$u$-channel exchange of selectrons, wherein the neutralinos are crossed
in the final state.}
\label{fig:radneutralino}
\end{figure}

%______________________________________________
In this Section We shall  calculate the cross section for the  radiative 
neutralino production process 
\begin{eqnarray}
e^-(p_1) + e^+(p_2) \rightarrow \tilde{\chi}_1^0(k_1) + \tilde{\chi}_1^0(k_2)
+ \gamma(q),
\label{radiative1}
\end{eqnarray}
in $SU(5)$ and $SO(10)$ grand unified theories with nonuniversal gaugino
masses at the grand unified scale. The  
symbols in the brackets denote the four momenta of the corresponding  
particles. At the tree level, the Feynman diagrams contributing to 
the radiative neutralino production
process are  shown in Fig.~\ref{fig:radneutralino}. In order 
to calculate the cross section for the radiative
process (\ref{radiative1}), 
we require the couplings of the neutralinos to electrons, 
the selectrons, and to the $Z^0$ bosons. 
These couplings are obtained from the 
neutralino mixing matrix (\ref{mssmneut}) as 
in  the Appendix~\ref{neut mass mat},
with the values of the soft SUSY breaking
gaugino mass parameters $M_1$ and $M_2$ for the 
respective grand unified theory,  
as calculated in Section~\ref{sec:gaugino mass patterns}.

The elements of the neutralino mixing matrix $N_{1j}$
for the case of $SU(5)$ and $SO(10)$, which are
relevant for our calculations, were calculated in the previous section.
As indicated earlier, as
a benchmark, we shall  calculate the radiative neutralino 
cross section for the MSSM with universal boundary condition~(\ref{gauginogut})
for the gaugino mass parameters at the GUT scale, for which we will
work with the parameters in the  MSSM electroweak
symmetry breaking scenario~(EWSB)~\cite{Pukhov:2004ca}.
This set of parameters  is summarized  in Table~\ref{parEWSB}. 
The relevant
elements of the neutralino mixing matrix  $N_{1j}$ are
summarized  in (\ref{ewsbcomp}). 

%%%%%%%%%%%%%%%%%%%%%%%%%%%%%%%%%%%%%%%%%%%%%%%%%%%%%%%%%%%%%%%%%%

%%%%%%%%%%%%%%%%%%%%%%%%%%%%%%%%%%%%%%%%%%%%%%%%%%%%%%%%%%%%%%%%%%%%%%%%
%__________________________________________________________

%%%%%%%%%% Feynman Diagrams for Radiative Neutralino Production 
%__________________________________________________________

\noindent
%%%%%%%%%%%%%%%%%%%%%%%%%%%%%%%%%%%%%%%%%%%%%%%%%%%%%%%%%%%%%%%%%%%%%%
%%%%% End of Feynman Diagrams for Radiative Neutralino Production %%%%
%%%%%%%%%%%%%%%%%%%%%%%%%%%%%%%%%%%%%%%%%%%%%%%%%%%%%%%%%%%%%%%%%%%%%%
%%%%%%%%%%%%%%%%%%%%%%%%%%%%%%%%%%%%%%%%%%%%%%%%%%%%%%%%%%%%%%%%%%%%%%%

\subsection{Cross Section for the Signal Process}
From Fig.~\ref{fig:radneutralino}, we see that 
the $t$- and $u$-channel exchange of right and left selectrons
$\tilde e_{R,L},$  and via $Z$ boson exchange in the $s$ channel
contribute to the process (\ref{radiative1}). 
The differential cross section for the process (\ref{radiative1}) 
can be written as~\cite{Grassie:1983kq, Eidelman:2004wy}
\begin{eqnarray}
d \sigma &=& \frac{1}{2} \frac{(2\pi)^4}{2 s}
\prod_f \frac{d^3 \mathbf{p}_f}{(2\pi)^3 2E_f}\delta^{(4)}(p_1 +
p_2 - k_1 - k_2 - q)|\M|^2,
\label{phasespace}
\end{eqnarray}
where $\mathbf{p}_f$ and $E_f$ are the final three-momenta
$\mathbf{k}_1$, $\mathbf{k}_2$, $\mathbf{q}$
and the final energies
$E_{\chi_1}$, $E_{\chi_2}$, and $E_\gamma$
of the neutralinos and the photon, respectively.
Using the standard technique, we average 
over initial spins and sum over the 
spins of the outgoing neutralinos. We also  sum over the polarizations
of the outgoing photon. 
Putting all this into effect,  the squared matrix element 
$|\M|^2$ in (\ref{phasespace})
can be written as~\cite{Grassie:1983kq}
\begin{eqnarray}
|\M|^2 & = & \sum_{i \leq j} T_{ij}, \label{squaredmatrix}
\end{eqnarray}
where $T_{ij}$ are squared amplitudes corresponding to the Feynman diagrams
in  Fig.~\ref{fig:radneutralino}. 
The phase space for the radiative neutralino production process 
in  (\ref{phasespace}) is described in detail in~\cite{Grassie:1983kq}.

%____________________________________________________________________________

\subsection{Radiative Corrections}
\label{sec:radiative corrections}

The next generation linear colliders are  designed to have
a high luminosity, which in turn will require beams with bunches 
of high densities. Due to the above requirement of high number density 
in the bunch, there arise problems
with the generation of strong electromagnetic fields in and around the bunches. This in turn
generates initial state radiation (ISR) and beamstrahlung effects. These effects 
have been  studied extensively in the past~\cite{Kureav, Nicorsini,Skrzypek:1990qs,Blankenbecler:1987rg}
in the context of the future linear colliders.
Therefore ISR and beamstrahlung effects have to be considered in any realistic calculation of the 
cross sections at a future linear collider since they result in the loss of beam energy
along with the disturbance of the initial beam calibration. 

The majority of the emitted photons
are soft and are lost down the beam pipe. Only the hard photons can be 
tagged. The radiated hard photon usually carries
from the radiating particle a certain amount of energy, resulting in 
the energy distribution  of the
initial beams. So a precise knowledge of ISR along with distribution of the 
photon spectrum resulting from beamstrahlung
and the behaviour of the electrons, positrons after the emission is required. 
We have calculated the radiative effects for our 
process and the background processes
using CalcHEP ~\cite{Pukhov:2004ca}, with parameters given in 
Table~\ref{ilc_par}~\cite{Brau:2007zza}. 
The resulting spectrum of electrons 
has been calculated using the structure function of the initial 
leptons valid upto 
all orders in perturbation theory. 
We note that the radiative effects are included in all our calculations
of the signal and background processes.

\begin{table}%[b]
\renewcommand{\arraystretch}{1.0}
\begin{center}
\vspace{0.5cm}
\begin{tabular}{||c|c||}
\hline 
Collider Parameters &ILC \\ \hline
$\sigma_x$ (nm) & 640  \\ 
 &\\
$\sigma_y$ (nm) & 5.7 \\  
&\\
$\sigma_z$ ($\mu$m) & 300 \\  
& \\
N ($10^{10}$) &2 \\  \hline
\end{tabular}
\end{center}
\caption{Beam parameters for the ILC, where N is the number of particles in the bunch and $\sigma_x$, $\sigma_y$
are the transverse bunch sizes at the interaction point, with $\sigma_z$ as the bunch length.}
\label{ilc_par}
\end{table}

%__________________________________________________________________________
\bigskip
\section{Numerical Results}
\label{sec:numerical}
The  tree-level cross section for radiative neutralino production
(\ref{radiative1}), and the standard model background from radiative 
neutrino and sneutrino production, (\ref{radiativenu}) and 
(\ref{radiativesnu}),  have  been calculated
using the program  CalcHEP~\cite{Pukhov:2004ca}. 
As noted above we have included the effects of radiative corrections
to the signal as well as the background processes.
We recall that the tree level cross sections have infrared and 
collinear divergences, 
which must be regularized by imposing cuts on the 
fraction of beam energy carried by the photon and the
scattering angle of the photon~\cite{Grassie:1983kq}. 
To implement this regularization, we define the fraction of the
beam energy carried by the photon as $ x = E_{\gamma}/E_{\rm beam},$
where  $ \sqrt {s} = 2E_{\rm beam}$ is the center of mass energy, and 
$E_{\gamma}$ is the energy carried away by the photon. 
We then impose the following cuts on $x$,  and on the scattering 
angle $\theta_{\gamma}$ of the photon~\cite{Dreiner:2006sb}:
\begin{eqnarray}
0.02 & \le & x \le  1-\frac{m_{\chi_1^0}^2}{E_{\rm beam}^2}, \label{cut1} \\
-0.95 & \le & \cos\theta_\gamma \le 0.95.  \label{cut2}
\end{eqnarray}
Note that the lower cut on $x$ in (\ref{cut1}) corresponds to photon energy 
$E_\gamma= 5$\, GeV for $\sqrt{s}=500$\, GeV. The
upper bound  of   $(1-m_{\chi_1^0}^2/E_{\rm beam}^2)$ on $x$  
corresponds to the kinematical limit of radiative neutralino production
process. The detector acceptance cut on the photon is applied so as to 
enhance the signal over the main irreducible background coming from the 
radiative neutrino production. We did not find any other cuts which would 
significantly reduce the background.

The  mass of the lightest neutralino for MSSM,  with universal boundary conditions
on the gaugino mass parameters at the GUT scale is taken to be
$m_{\chi_1^0}=108$~GeV from the EWSB scenario, to implement the 
cuts on the photon energy in the calculation of the cross sections.  
Using Eq.~(\ref{cut1}) we get
a fixed upper limit $E_\gamma^{\rm max} \simeq 203.3$~GeV for MSSM 
at $\sqrt{s}=500$~GeV for the photon energy. We have used  
these cuts for both signal and  background processes for all the scenarios
that we have considered in this paper.

% Two different procedures  have
% been chosen for the calculation of the cross section 
% for the production of neutrinos as the main
% background process~(\ref{radiativenu}), 
% one  ``with upper cut''and another ``without upper cut,''  
% for obvious reasons. 
% We note that at $\sqrt{s}=500$ GeV and for $m_{\chi_1^0} \gsim 70$ GeV,
% this cut reduces a substantial amount of the on-shell $Z$ boson 
% contribution to radiative neutrino production process.  

\begin{table}%[b]
\renewcommand{\arraystretch}{1.0}
\begin{center}
\vspace{0.5cm}
\begin{tabular}{||c|c|c|c|c||}
\hline 
$\sqrt{s}$ &$\sigma_{EWSB}$ &$\sigma_{EWSB} + R.E.$ &$\sigma_{SM}$  &$\sigma_{SM} + R.E.$ \\
& & & & \\ \hline
  GeV &(fb)$\times 10^{-1}$ &(fb)$\times 10^{-1}$ &(fb)$\times 10^3$ &(fb)$\times 10^3$ \\
 & & & & \\  \hline
300          &0.9072  &0.7586  &2.1899  &2.4187 \\  
 & & & & \\
400          & 1.4035  &1.2821 &2.3691  &2.4373 \\  
& & & & \\
500          & 1.3963  &1.3377 &2.4191  &2.4321 \\ 
& & & & \\
600          & 1.2540  &1.2383 &2.4329  &2.4518 \\ 
& & & & \\
700          & 1.0922  &1.1030 &2.4226  & 2.4359 \\
& & & & \\
800          & 0.9443   &0.9721 &2.3980  &2.3648 \\ 
& & & & \\
900          & 0.8179   &0.8568 &2.3687  &2.3318 \\ 
& & & & \\
1000         & 0.7124   &0.7576 &2.3341  &2.2934 \\  \hline

\end{tabular}
\end{center}
\caption{Cross section of the signal process in MSSM EWSB scenario, along with the the SM irreducible background 
with and without the inclusion of the radiative effects~(R.E.) due to ISR and beamstrahlung.}
\label{isr_EWSB}
\end{table}

%%%%%%%%%%%%%%%%%%%%%%%%%%%%%%%%%%%%%%%%%%%%%%%%%%%%%%%%%%%%%%
\begin{figure}[ht]
\begin{minipage}[b]{0.45\linewidth}
\vspace*{0.45cm}
\centering
\includegraphics[width=6.5cm, height=5cm]{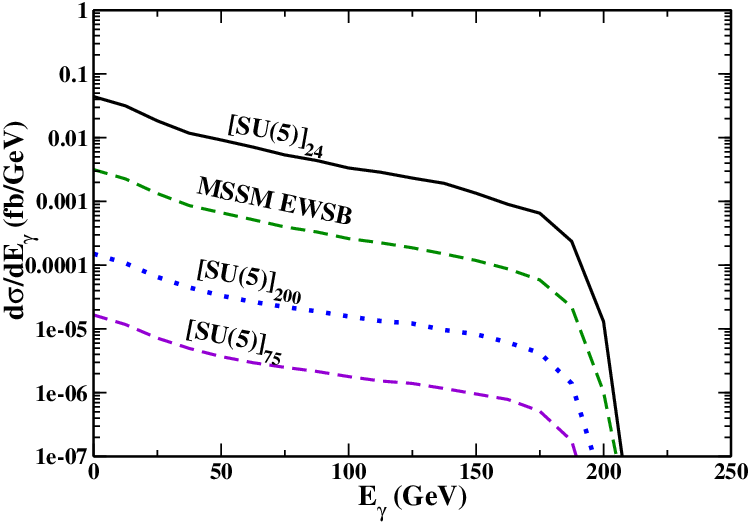}
\caption{The photon energy distribution 
$d\sigma/dE_{\gamma}$
for the radiative neutralino production including radiative effects for  the $SU(5)$ 
grand unified theory with nonuniversal gaugino masses
compared with MSSM EWSB with universal 
gaugino masses.}
\label{fig:SU(5)}
\end{minipage}
\hspace{0.7cm}
\begin{minipage}[b]{0.45\linewidth}
\centering
\includegraphics[width=6.5cm, height=5cm]{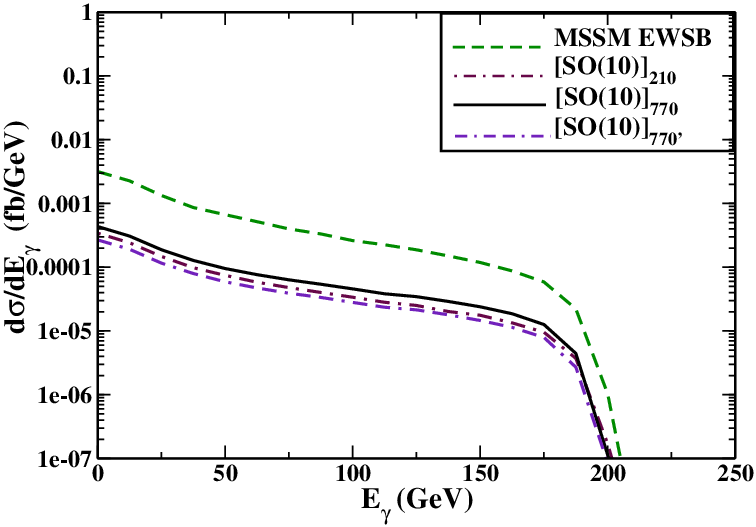}
\caption{The photon energy distribution 
$d\sigma/dE_{\gamma}$ including radiative effects
for the radiative neutralino production for  $SO(10)$
grand unified theory with nonuniversal gaugino masses 
compared with MSSM EWSB with universal gaugino masses.}
\label{fig:SO(10)}
\end{minipage}
\end{figure}
    
%________
\subsection{Photon Energy~($E_\gamma$) Distribution and Total Beam
Energy~($\sqrt {s}$) Dependence}
\label{sec:photon_energ_distribution}
To begin with, we have calculated the energy distribution of the photons 
from radiative neutralino production in MSSM EWSB scenario as well as for
the different GUT models considered in this paper.
%_____________________________________

%_________________________________________

\begin{figure}[htb]
\begin{minipage}[b]{0.45\linewidth}
\centering
\vspace*{0.45cm} 
\includegraphics[width=6.5cm, height=4cm]{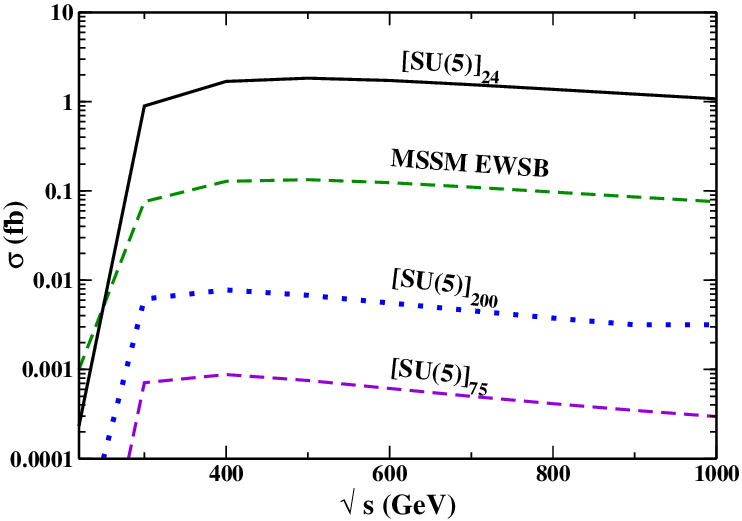}
\caption{Total cross section  $\sigma$ for the radiative process
$e^+e^- \to \tilde\chi^0_1 \tilde\chi^0_1\gamma$, with the inclusion of radiative effects
as a function of center of mass energy $\sqrt s$
for the  $SU(5)$ grand unified theory
with nonuniversal gaugino masses  compared with
MSSM EWSB scenario with universal gaugino masses at the grand unified scale.}
\label{fig:SU(5)_cs}
\end{minipage}
\hspace{0.7cm}
\begin{minipage}[b]{0.45\linewidth}
\centering
\includegraphics[width=6.5cm, height=4cm]{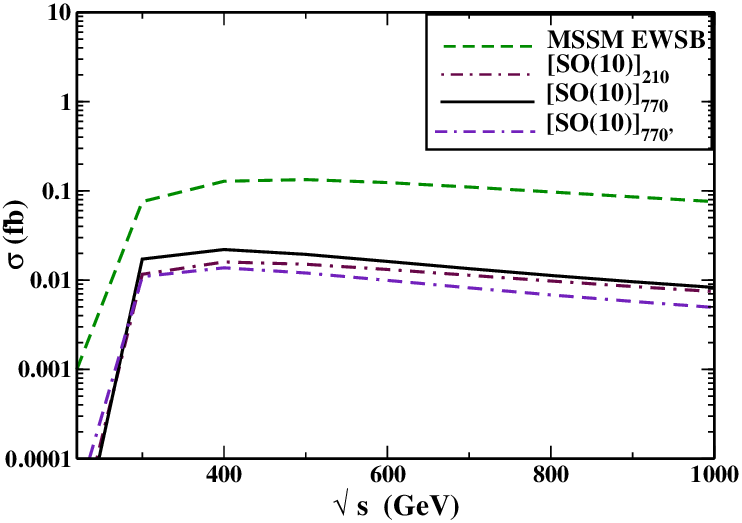}
\caption{Total cross section  $\sigma$ for the radiative process
$e^+e^- \to \tilde\chi^0_1 \tilde\chi^0_1\gamma$, with the inclusion of radiative effects
as a function of center of mass energy $\sqrt s$ for the $SO(10)$
grand unified theory with nonuniversal gaugino masses
compared with
MSSM EWSB scenario with universal gaugino masses at the grand unified scale.}
\label{fig:SO(10)_cs}
\end{minipage}
\end{figure} 
%___________________________________________________

In Table~\ref{isr_EWSB} we illustrate the  changes in the signal cross section 
with and without the inclusion of radiative effects for the MSSM EWSB scenario, along with the irreducible SM background
with the application of cuts described earlier. 
The radiative effect leading to the distribution of the beam energy 
can be distinctly seen from this Table, with the increase of cross section 
at higher c.m. energies and the corresponding  decrease at lower c.m. energies
for the signal process. Similar behaviour holds in other models for the 
signal process and the supersymmetric background processes. In case of the 
background process, the radiative neutrino production,
the presence of the $Z$ pole along with the massless 
particles in the final state leads to a decrease of cross section at 
higher c.m. energies and vice versa. 
The plots here  only show the results with the inclusion of radiative 
effects of ISR and beamstrahlung, since the behaviour in case of 
different scenarios for the signal process and supersymmetric background
with and without the inclusion of radiative effects is similar to the MSSM EWSB scenario shown in Table~\ref{isr_EWSB}.

In Figs.~\ref{fig:SU(5)} and \ref{fig:SO(10)} we  show
the energy distribution of photons for models
with nonuniversal gaugino masses in grand unified 
theories based on  $SU(5)$ and $SO(10)$, where we have
included the radiative effects. We have compared this 
with the photon energy distribution for the MSSM EWSB model with 
universal gaugino masses at the GUT scale.  The  energy dependence
of the total cross section for these models is also calculated and this
is shown in
Figs.~\ref{fig:SU(5)_cs} and \ref{fig:SO(10)_cs}. From Figs.~\ref{fig:SU(5)} 
and \ref{fig:SO(10)}
as well as  from Figs.~\ref{fig:SU(5)_cs} and \ref{fig:SO(10)_cs}, 
it can be seen that
the signal in case of MSSM EWSB and $[SU(5)]_{24}$ is enhanced compared to 
the other scenarios considered here. This is mainly due to the fact 
that the dominant component of the neutralino
in $[SU(5)]_{24}$ is a bino, whereas in other cases 
the lightest neutralino is dominantly a higgsino state. The MSSM EWSB scenario predicts
a lightest neutralino with a dominant higgsino component,  but it also has a significant
bino component leading to the enhancement of right selectron-electron-neutralino coupling.
For the other cases with a higgsino like neutralino the $t$- and $u$- channel exchange of 
$\tilde{e}_{R,L}$, is suppressed, 
with the only contribution coming from off
shell $Z$ decay. 
In order to find ways of enhancing the signal, 
one must study the dependence of the signal  on selectron $\tilde{e}_{R,L}$ masses as well as
on the parameters $\mu$ and $M_2$, which determine the neutralino mixing elements.

%%%%%%%%%%%%%%%%%%%%%%%%%%%%%%%%%%%%%%%%%%%%%%%%%%%%%%%%%%%%%%%%%
\subsection{Dependence on $\mu$ and $M_2$}

As can be seen from the neutralino mass matrix in the Appendix~\ref{exp_cons},
the mass of the lightest neutralino depends 
on the parameters $\mu$ and $M_2$. Therefore,  it is important to
study the dependence of cross section of the signal process
on these parameters.  Since  $\mu$ and $M_2$ are independent
parameters, we have studied  the dependence of the cross section
$\sigma$($e^+e^- \to \tilde\chi^0_1 \tilde\chi^0_1\gamma$) 
on these parameters independently.  We have considered here 
all the scenarios with both universal and nonuniversal gaugino 
masses. The values of the parameters $\mu$ and $M_2$ are  chosen 
so as to avoid color and charge breaking minima, unbounded from below
constraint on scalar potential, and to satisfy phenomenological constraints 
on different sparticle masses discussed in Appendix~\ref{exp_cons}. 

We note here that we have carried out a check on the parameter space used
in our calculations as to whether the complete scalar potential has charge
and color breaking minima~(CCB) which is lower than the electroweak minimum.
We have also checked whether the scalar potential is unbounded 
from below~(UFB). The criteria used for these conditions are
\bea
A_f^2 & < & 3(m_{\tilde f_L}^2 + (m_{\tilde f_R}^2 + \mu^2 + m_{H_2}),
\label{ccb1}\\
m_{H_2} +  m_{H_1} & \ge & 2|B\mu|,
\label{ufb1}
\eea
respectively, at a scale $Q^2 > M^2_{\rm EWSB}$. Here $f$ denotes the
fermion generation, and A is the trilinear supersymmetry breaking parameter.
We have implemented these conditions through the 
SuSpect package~\cite{Djouadi:2002ze} which computes
the masses and couplings of the supersymmetric partners of the SM particles.
For each model considered in this paper, we perform the renormalization group
evolution to calculate the particle spectrum. While doing so we check 
for the consistency of the chosen parameter set
with electroweak symmetry breaking and that the conditions (\ref{ccb1}) and
(\ref{ufb1}) are satisfied.

\begin{figure}[htb]
\begin{minipage}[b]{0.45\linewidth}
\centering
\vspace*{0.7cm}
\includegraphics[width=6.5cm, height=5cm]{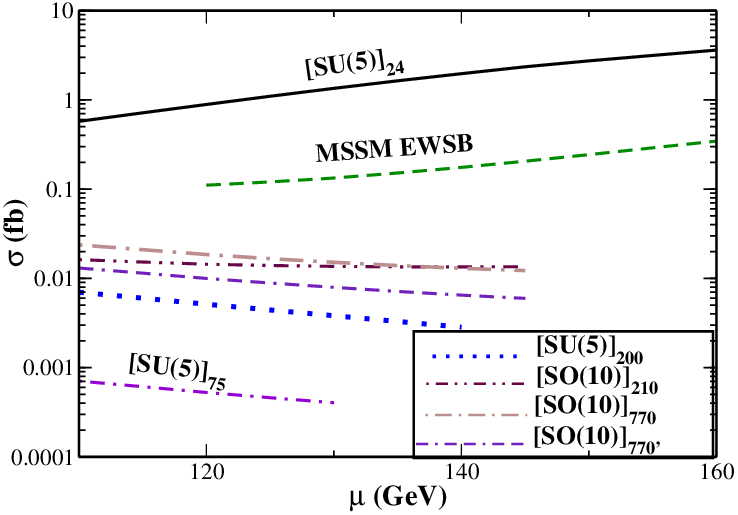}
\caption{The total  radiative neutralino production cross section $\sigma$ with radiative effects included
as a function of $\mu$  in the range $\mu$ {\Large{$\epsilon$}} [110, 160] GeV
for different models considered in this paper at $\sqrt{s} = 500$~GeV.}
\label{fig:mu}
\end{minipage}
\hspace{0.4cm}
\begin{minipage}[b]{0.45\linewidth}
\centering
\includegraphics[width=6.5cm, height=5cm]{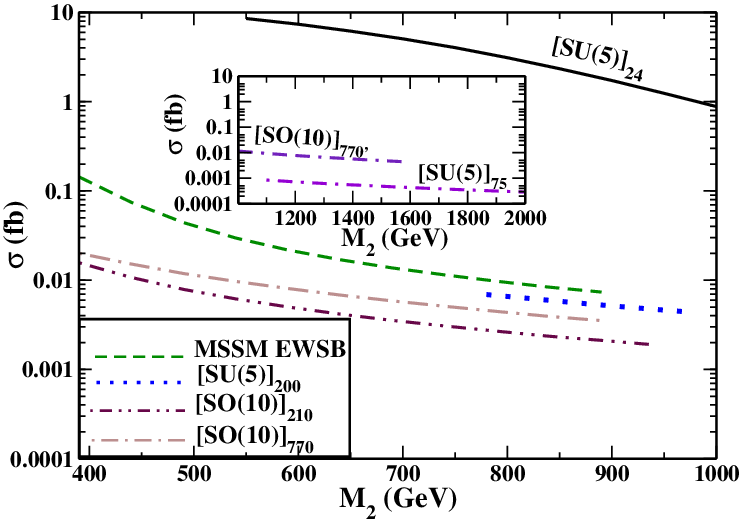}
\caption{Total cross section $\sigma$ with the inclusion of radiative effects for the radiative neutralino
production as a function of  $M_2$ for different models with
$M_2$ {\Large{$\epsilon$}} [390, 1000] GeV
at $\sqrt{s} = 500$~GeV.}
\label{fig:M_2dep}
\end{minipage}
\end{figure}

In Fig.~\ref{fig:mu} we show the $\mu$ dependence of the cross section for
different models considered in this paper. 
For a wide range of values of $\mu$, the $[SU(5)]_{24}$ scenario
satisfies  all experimental constraints, with $\tilde{\chi}_1^0$ as the LSP.
Since the neutralino in this case is mainly a bino like state, with values of 
$\mu$ $> M_1$, the  neutralino mass is relatively insensitive to the 
values of $\mu$.
For the other scenarios with a higgsino type lightest neutralino, the cross 
section is sensitive to the value of $\mu$. 
Since $m_{\tilde{\chi}_1^0} \propto \mu$, above  a certain value of 
$\mu$, $\tilde{\chi}_1^0$ ceases to be the lightest supersymmetric
particle. Depending on the percentage  of the higgsino component, 
the cross section changes  with value of $\mu$. In $[SU(5)]_{75}$,
since $M_1, M_2 \gg \mu$,  $m_{\tilde{\chi}_1^0}$  depends on $\mu$, 
so from Fig.~\ref{fig:mu}, it can be seen that the signal cross section decreases 
with $\mu$ due to increasing $m_{\tilde{\chi}_1^0}$.
Most of the scenarios considered here are tightly constrained  
as a function of $\mu$, with neutralino as
the  LSP. This is due to the various limits  on the sparticles masses from the 
experiments. The cross section for some 
scenarios in this region  is too small to be observed  at ILC with  
$\sqrt{s} = 500$~GeV, even with an integrated luminosity of $500$ fb$^{-1}$.

%%%%%%%%%%%%%%%%%%%%%%%%%%%%%%%%%%%%%%%%%%%%%%%

We next show in Fig.~\ref{fig:M_2dep} the dependence of the radiative neutralino cross section
on the soft gaugino mass parameter $M_2$ for different models that we have 
studied in this paper. From this  Fig. we note that
the total cross section decreases with increasing values of $M_2$. 
The elements of the neutralino mixing matrix changes with the variation
of the wino parameter $M_2$, which in turn change the values of the vertices  
contributing to  the radiative neutralino production, Table~\ref{feynmandiag}
in the Appendix~\ref{neut mass mat}. The range of $M_2$
considered for the different models is taken from Fig.~\ref{fig:m1m2}. The variation of the
cross section with $M_2$ for $SU(5)_{75}$ and $SO(10)_{770'}$ are shown in the inset of
Fig.~\ref{fig:M_2dep}, as their $M_2$ range is different from the other scenarios.
A lower value of $M_2$ favours a cross section which can
be measured experimentally for some scenarios, whereas for higher values 
the difficulty in measurement of cross section increases.

\subsection{Dependence on selectron masses}

\begin{figure}[t]
\begin{minipage}[b]{0.45\linewidth}
\centering
\vspace*{0.7cm}
\psfrag{L}{$\bf{m_{\tilde{e}_L}} $}
\includegraphics[width=6.5cm, height=5cm]{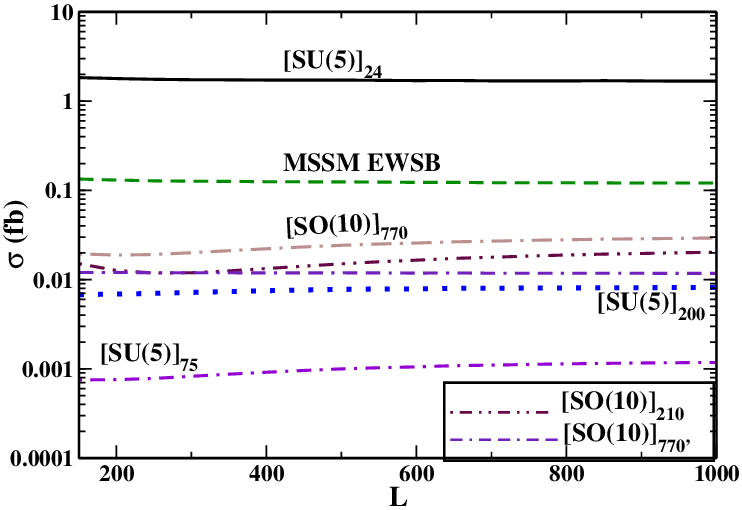}
\caption{Total cross section $\sigma$ for the radiative neutralino production with radiative effects included versus
$m_{\tilde{e}_L}$  at $\sqrt{s} = 500 $~GeV.}
\label{fig:sneu_lt}
\end{minipage}
\hspace{0.4cm}
\begin{minipage}[b]{0.45\linewidth}
\centering
\psfrag{R}{$\bf{m_{\tilde{e}_R}} $}
\includegraphics[width=6.5cm, height=5cm]{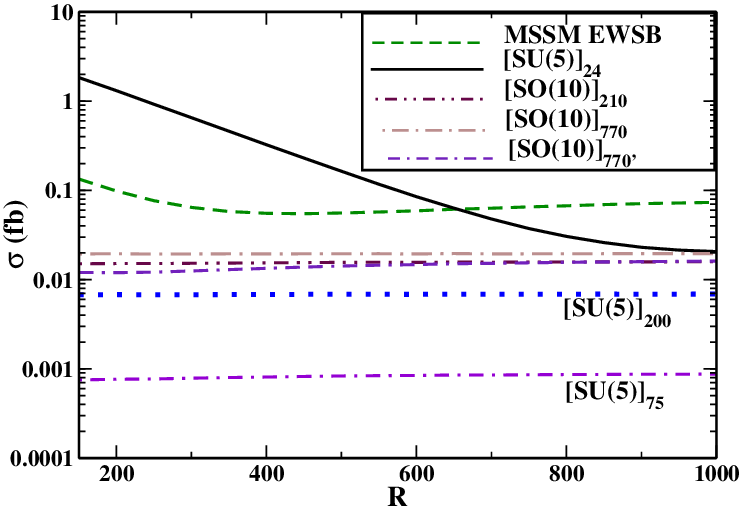}
\caption{Total cross section $\sigma$ along with radiative effects for the radiative neutralino production versus
$m_{\tilde{e}_R}$  at $\sqrt{s} = 500 $~GeV.}
\label{fig:sneu_rt}
\end{minipage}
\end{figure}

The signal process $\sigma(e^+e^-
\to\tilde\chi^0_1\tilde\chi^0_1\gamma)$  mainly proceeds via right and left
selectron $\tilde e_{R,L}$ exchange in the $t$- and $u$-channels. 
Different models  that we have considered in this work have the selectron 
masses as independent parameters. 
Figs.~\ref{fig:sneu_lt} and  \ref{fig:sneu_rt}  show the dependence of the
total cross section $\sigma$ for the  radiative
neutralino production on the left and right selectron masses. 
From Fig.~\ref{fig:sneu_lt} we see that the cross section is not
sensitive to the left selectron mass for $[SU(5)]_{24}$
scenario, because the LSP is predominantly a bino.
As  a consequence, the dominant
contribution to the cross section comes from the right selectron exchange,
with the contribution from the left selectron exchange suppressed by
$\tan\theta_W$ in the coupling, as can be seen from  Table~\ref{feynmandiag}
in the Appendix~\ref{neut mass mat}. As a consequence the cross section 
has very little sensitivity 
to the left selectron mass, but is sensitive to the right selectron mass in the range $150$ - $1000$ GeV. 
The MSSM EWSB scenario has a LSP with a dominant contribution coming from 
higgsino,
but has a significant bino contribution as well. Due to this the cross 
section is relatively
insensitive to mass of $\tilde e_{L}$, but a sensitivity is seen 
with respect to right selectron mass.
  
In the other scenarios with a higgsino type LSP, the neutralino
coupling with $\tilde e_{R,L}$
is suppressed, making these almost insensitive to the selectron masses. 
As can be seen from the 
Figs.~\ref{fig:sneu_lt} and  \ref{fig:sneu_rt}, the cross section 
for these scenarios has very little sensitivity 
to the left and right selectron mass. The suppressed couplings lead to a 
reduction in the cross section.

%___________________________________  

%__________________________________________________________________________
%\bigskip
%___________________________________________________________________
%  The background processes begin here 
%__________________________________________________________________
\section {Background Processes}
\label{sec:backgrounds}
\subsection{The Neutrino Background}\label{sec:neu}
\noindent

The SM radiative neutrino production is the main irreducible background
for the process radiative neutralino production~(\ref{radiative1}).  The other possible backgrounds
are from $e^+e^- \rightarrow \tau^+ \tau^- \gamma$, with both the $\tau's$ decaying to soft leptons or hadrons
but the contribution from this process is found to be negligible.  Another large background 
is from the radiative Bhabha  scattering, 
$ e^+e^- \rightarrow  e^+ e^- \gamma$, where  ${e^\pm}'s$ are not
detected.  This  radiative scattering is usually eliminated by imposing 
a cut on $E_\gamma$. The 
events are selected by imposing the condition that any particle other 
than $\gamma$ appearing in the
angular range $-0.95 < \cos \theta_\gamma < 0.95$ must have energy less than 
$E_{max}$, where $E_{max}$ is detector dependent, but presumably
no larger than a few GeV. This is discussed in detail in 
the literature~\cite{Chen}.

\begin{figure}[htb]
\includegraphics{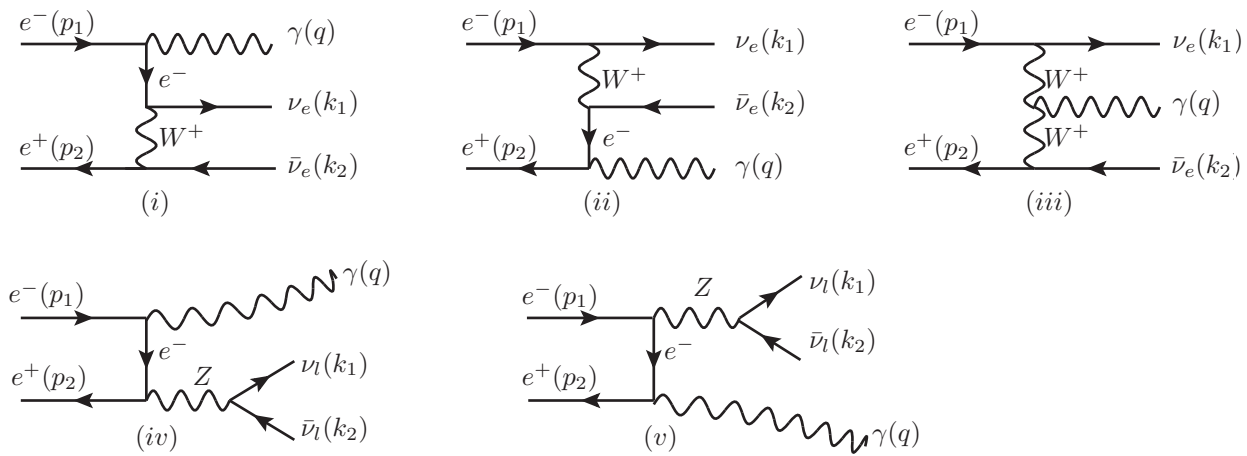}
\caption{Feynman diagrams contributing to the radiative neutrino process 
 $e^+e^- \rightarrow {\nu}{\bar{\nu}}\gamma$ where ($iv$ and $v$)
 corresponds to the neutrinos of three flavours}
\label{fig:radneutrino}
\end{figure}

The SM radiative neutrino production 
\begin{equation}
e^+ +e^- \to \nu_\ell+\bar\nu_\ell+\gamma\,,\;\;\qquad \ell=e,\mu,\tau,
\label{radiative2}
\end{equation}
has been studied extensively~\cite{Datta:1996ur,Gaemers:1978fe,Berends:1987zz,
Boudjema:1996qg,Montagna:1998ce}.
For this background process $\nu_e$ are produced via
$t$-channel $W$ boson exchange, and $\nu_{e,\mu,\tau}$ via $s$-channel
$Z$ boson exchange.  The corresponding Feynman diagrams are shown in
Fig.~\ref{fig:radneutrino}.

%_____________

%%%%%%%% The neutrino diagrams begin here. %%%%%%%%%%%%%%%%%%%%%%%%%%
%%%%%%%%%%%%%%%%%%%%%%%%%%%%%%%%
\noindent
%%%%%%%%%%%%%%%%%%%%%%%%%%%%%%%%%%%
% The neutrino diagrams end here. %
%%%%%%%%%%%%%%%%%%%%%%%%%%%%%%%%%%%

\begin{figure}[htb]
\begin{minipage}[b]{0.4\linewidth}
\centering
\vspace*{0.7cm}
\includegraphics[width=6.5cm, height=5cm]{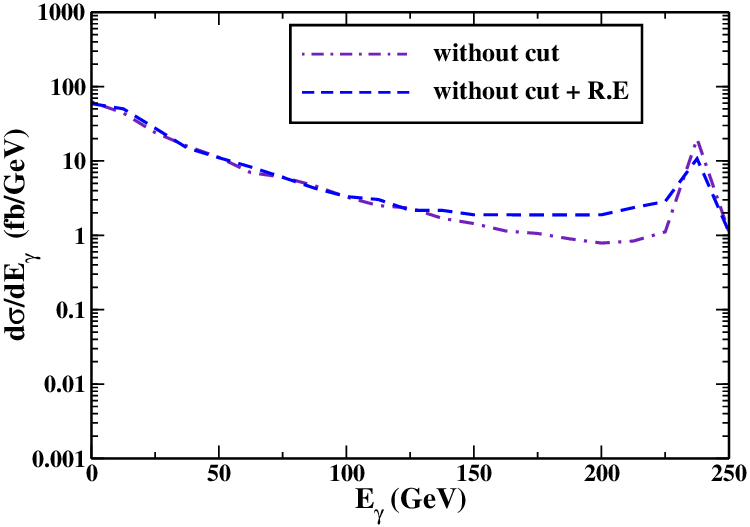}
\caption{Plot showing the photon energy distribution 
$d\sigma/dE_{\gamma}$
for the radiative neutrino production process
$e^+e^- \rightarrow \nu \bar\nu \gamma$ at $\sqrt{s} = 500$~GeV, with and without including radiative effects (R.E.).}
\label{fig:neu_pe}
\end{minipage}
\hspace{0.7cm}
\begin{minipage}[b]{0.4\linewidth}
\centering
\includegraphics[width=6.5cm, height=5cm]{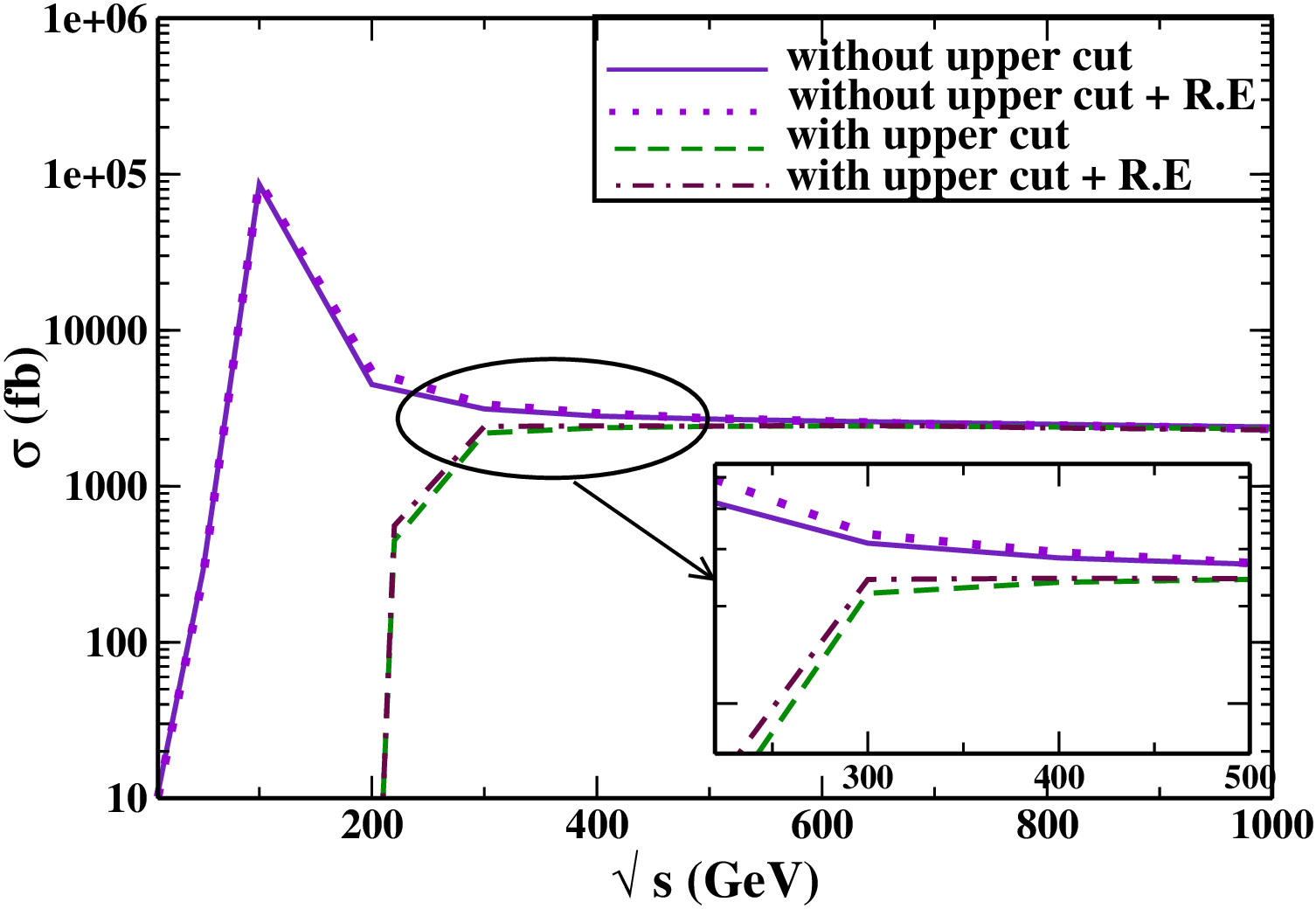}
\caption{The total energy $\sqrt{s}$ dependence of the  radiative neutrino 
cross section with  and without an 
upper cut  on the photon energy $E_{\gamma}$, along with and without the consideration of radiative effects (R.E.).}
\label{fig:neu_cs}
\end{minipage}
\end{figure}

The background photons from this process tend to be mostly in the 
forward and backward directions as compared to the signal photons, 
therefore the 
detector acceptance cut for the photon is applied as $|\cos\theta_\gamma|<0.95$.
This SM background has the same  photon energy distribution $d\sigma/d E_\gamma$
and the $\sqrt s$ dependence of the cross
section $\sigma$ for all the scenarios studied in this paper.
Fig.~\ref{fig:neu_pe} shows that the photon energy distribution from 
the radiative neutrino production peaks at
$E_\gamma= (s -m_{Z^0}^2)/(2\sqrt{s}) \approx 244$~GeV because of  
the radiative $Z^0$ production($\sqrt s > m_{Z^0}$). The result is presented 
here with and without the higher order QED radiative  effects. The inclusion of
radiative effects make the peak due to the radiative return of $Z^0$ slightly broad.
By imposing an upper cut on the photon energy 
$x^{\rm max}=E_\gamma^{\rm max}/E_{\rm beam}=1-m_{\chi_1^0}^2
/E_{\rm beam}^2$, see Eq.~(\ref{cut1}), the photon background
from radiative neutrino production can be reduced, 
by elimination of the on-shell $Z^0$
contribution to  the background cross section.  
From the Feynman diagrams for 
the background process, Fig.~\ref{fig:radneutrino}, it can be seen
that this process has a strong polarization dependence on the initial beams due to the exchange of $W$ bosons
which couples only to left handed electron and right handed positron. 
Therefore, with a suitable choice of beam polarization
along with various kinematical cuts discussed earlier,
the contribution from the background process
can be significantly reduced.

    In  Fig.~\ref{fig:neu_cs} we show the $\sqrt s$ dependence of the
total radiative neutrino cross section, with and without the inclusion of higher order
QED radiative effects. 
Without the upper cut on the 
photon energy $x^{\rm  max}$, the background cross section from
radiative neutrino production $e^+e^- \to \nu\bar\nu\gamma$~ 
 is much larger than the
corresponding cross section with the cut, near the $Z^0$
production threshold. The main purpose of the cut is to move
away from the $Z^0$ peak.
When we impose the cut, the signal cross section from radiative
neutralino production is approximately three orders of magnitude smaller than 
the background in the case of MSSM EWSB and the various GUT models.

%_________________________________________________
%______________________________________________
%%%%%%%% The sneutrino diagrams start here. 
%______________________________________________
\begin{figure}[htb]
{%
\unitlength=1.0pt
\includegraphics{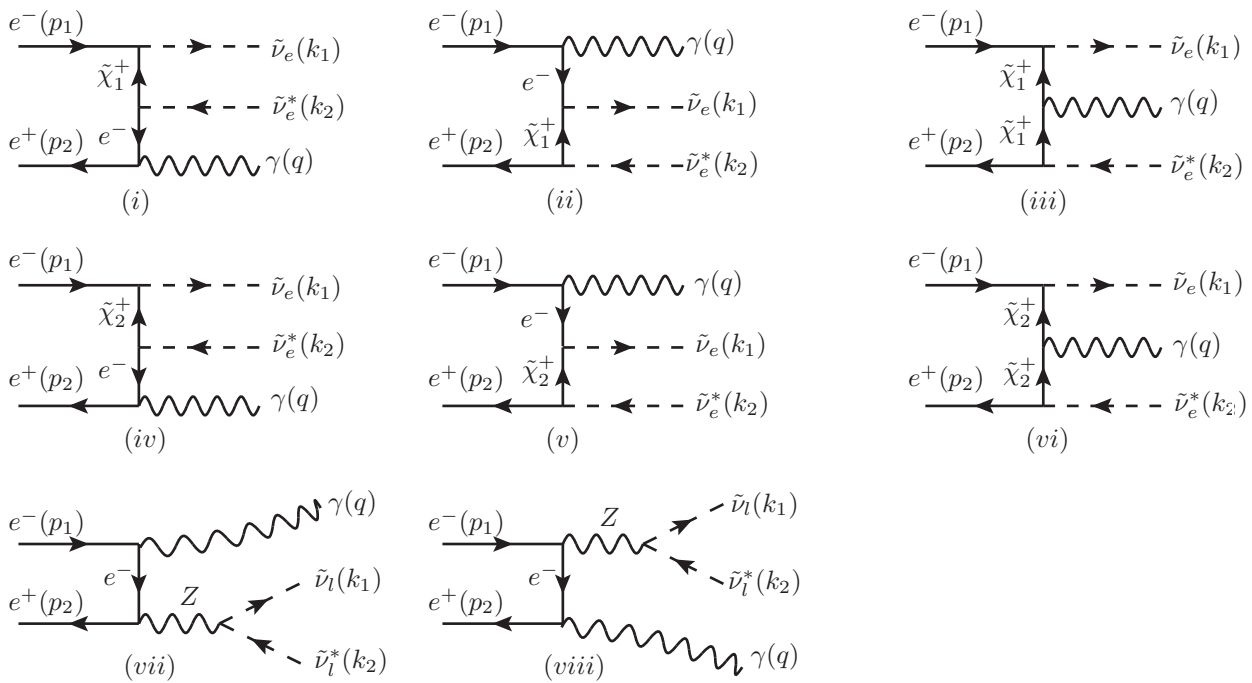}
}
\caption{Feynman diagrams contributing  to the radiative sneutrino production
process $e^+e^- \rightarrow \tilde{\nu}\tilde{\nu}^*\gamma$, with the last 
two diagrams ($vii$ and $viii$) corresponding to all the leptonic sneutrino }
\label{fig:radsneutrino}
\end{figure}
\noindent
%%%%%%%%%%%%%%%%%%%%%%%%%%%%%%%%%%%%%%%%%%%%%%%%%%%%%%%%%%%%%%%%%%%
%%%%% End of sneutrino diagrams %%%%%%%%%%%%%%%%%%%%%%%%%%%%%%%%%%%
%%%%%%%%%%%%%%%%%%%%%%%%%%%%%%%%%%%%%%%%%%%%%%%%%%%%%%%%%%%%%%%%%%%

\subsection{The Supersymmetric Background}
\begin{figure}[htb]
\begin{minipage}[b]{0.4\linewidth}
\centering
\vspace*{0.7cm}
\includegraphics[width=6.5cm, height=5cm]{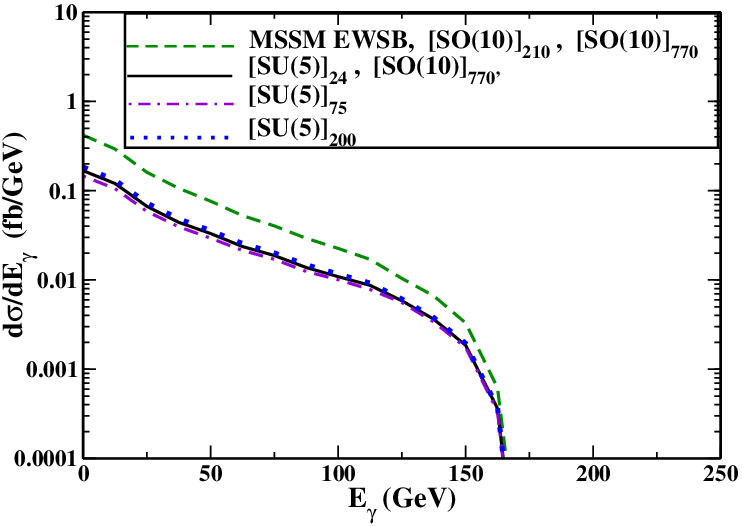}
\caption{Plot showing the photon energy distribution 
$d\sigma/dE_{\gamma}$
for the radiative sneutrino production process 
$e^+e^- \rightarrow \tilde{\nu}\tilde{\nu}^*\gamma$ at $\sqrt{s} = 500$~GeV, with the inclusion of radiative effects.}
\label{fig:sneu_pe}
\end{minipage}
\hspace{0.7cm}
\begin{minipage}[b]{0.4\linewidth}
\centering
\includegraphics[width=6.5cm, height=5cm]{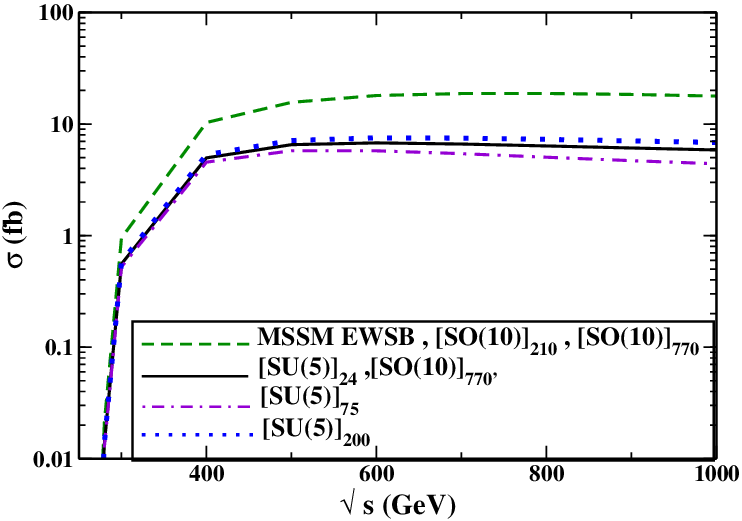}
\caption{The total energy $\sqrt{s}$  dependence of the radiative sneutrino 
cross section $e^+e^- \rightarrow \tilde{\nu}\tilde{\nu}^*\gamma$
with  an upper cut on the photon energy $E_\gamma$ and the inclusion of radiative effects.}
\label{fig:sneu_cs}
\end{minipage}
\end{figure}

The radiative
neutralino production  (\ref{radiative1}) has also a  supersymmetric background coming from the
sneutrino production process~\cite{Datta:1996ur, Franke:1994ph}

\begin{equation}
e^+ +e^- \to \tilde\nu_\ell+\tilde\nu^\ast_\ell+\gamma\,, 
\;\qquad \ell=e,\mu,\tau\,.
\label{radiative3}
\end{equation}
This is in addition to the  background  from the SM process
(\ref{radiative2}).
The lowest order Feynman diagrams contributing to the process 
(\ref{radiative3}) are shown in Fig.~\ref{fig:radsneutrino}.
This background process receives $t$-channel contributions via
virtual charginos for $\tilde\nu_e \tilde\nu_e^\ast $ production, as
well as $s$-channel contributions from $Z$ boson exchange for
$\tilde\nu_{e, \mu,\tau} \tilde\nu_{e, \mu,\tau}^\ast $ production. 
In Fig.~\ref{fig:sneu_pe} we show the photon energy 
distribution $d\sigma/d E_\gamma$ for radiative  sneutrino  production
$e^+e^- \to \tilde\nu\tilde\nu^\ast\gamma$ at $\sqrt{s} = 500$~GeV
for the MSSM EWSB and  grand unified scenarios. 
The total cross section 
for the radiative sneutrino production is shown in Fig.~\ref{fig:sneu_cs}.
Since $\tilde\nu_e \tilde\nu_e^\ast $ production receives $t$-channel 
contributions
via virtual charginos, the production cross section as well as the photon 
energy distribution depends on the chargino mixing matrix $\textbf{U}$ for the different scenarios.
The cuts applied to this process are the same as discussed before.

         As can be seen from Fig.~\ref{fig:sneu_cs} radiative sneutrino
production (\ref{radiative3}) can be a major supersymmetric 
background to neutralino production (\ref{radiative1}) if  sneutrinos
decay invisibly, e.g.  via $\tilde\nu\to\tilde \chi^0_1\nu$.
This scenario has been called ``virtual LSP'' scenario~\cite{Datta:1996ur}.  
However, if kinematically allowed, other visible decay channels 
like $\tilde\nu\to\tilde\chi^\pm_1\ell^\mp$ reduce the background rate 
from radiative sneutrino production.  In the 
scenarios with a bino type neutralino, 
as in  $[SU(5)]_{24}$,
the branching ratio ${\rm  BR}(\tilde\nu_e\to\tilde\chi_1^0\nu_e)=100\%$,
and, therefore, this process serves as a dominant background.
However, for the other cases where higgsino is the dominant component
we have the branching ratios as given in Table~\ref{br_neu}. The second lightest neutralino
being heavier than the sneutrino $\tilde\nu_e$ in MSSM EWSB scenario, the branching ratio is kinematically not accessible.

\begin{table}[htb]
\renewcommand{\arraystretch}{1.0}
\begin{center}
\vspace{0.5cm}
        \begin{tabular}{|c|c|c|c|c|c|c|}
\hline
Branching Ratios &MSSM EWSB&$SU(5)_{75}$ &$SU(5)_{200}$ &$SO(10)_{210}$ &$SO(10)_{770}$ &$SO(10)_{770'}$\\ \hline \hline
${\rm  BR}(\tilde\nu_e\to\tilde\chi_1^0\nu_e)$ &78.4\%&8.1\% &21.2\%   &18\%   &24.2\%  &44.4\% \\ \hline
${\rm  BR}(\tilde\nu_e\to\tilde\chi_2^0\nu_e)$ &      &1.8\%  &4.54\%   &0.8\%   &1.2\%  &6\%   \\ \hline
${\rm  BR}(\tilde\nu\to\tilde\chi^\pm_1\ell^\mp)$&21.6\% &90.1\%  &74.3\%  &81\% &74.8\% &49.6\%   \\ \hline
\end{tabular}
\end{center}
\vspace{-0.5cm}
\caption{Branching ratios of the sneutrino for different models with a higgsino type lightest neutralino}
\renewcommand{\arraystretch}{1.0}
\label{br_neu}
\end{table}

Furthermore, neutralino production $e^+e^- \to \tilde\chi_1^0
\tilde\chi^0_2$ followed by subsequent radiative neutralino
decay~\cite{Haber:1988px} $\tilde\chi^0_2 \to \tilde\chi^0_1 \gamma$
is also a potential background.  However, significant branching ratios
${\rm BR}(\tilde\chi^0_2 \to \tilde\chi^0_1 \gamma)>10\%$ are only
obtained for small values of $\tan\beta<5$ or $M_1\sim
M_2$~\cite{Ambrosanio:1995it,Ambrosanio:1995az,Ambrosanio:1996gz}.
Thus,  we neglect this background, detailed  discussions
of which  can be found in Refs.~\cite{Ambrosanio:1995az,Ambrosanio:1996gz,
Baer:2002kv}.
%______________________________________

%________________________________________________________
\subsection{Theoretical Significance}
\begin{figure}[htb]
\begin{minipage}[b]{0.4\linewidth}
\centering
\vspace*{0.7cm}
\includegraphics[width=6.5cm, height=5cm]{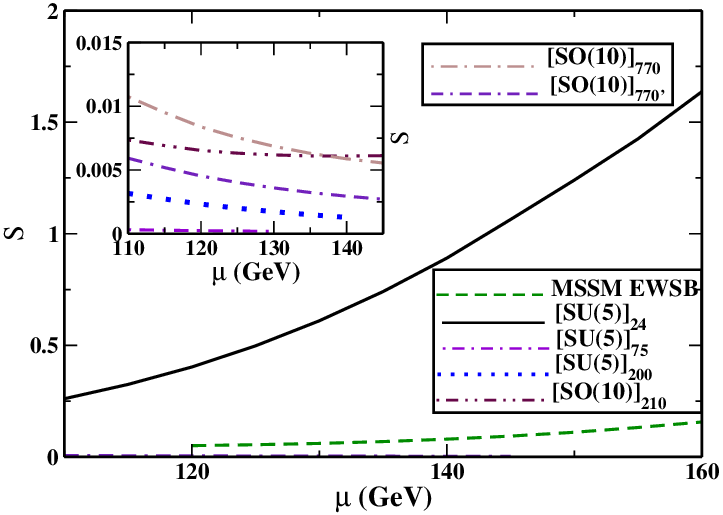}
\caption{Plot showing the theoretical significance 
$S$ for the radiative neutralino production as a function of  $\mu$ for 
different models considered in this paper with $\sqrt{s}= 500$~GeV.}
\label{fig:sig_mu}
\end{minipage}
\hspace{0.7cm}
\begin{minipage}[b]{0.4\linewidth}
\centering
\includegraphics[width=6.5cm, height=5cm]{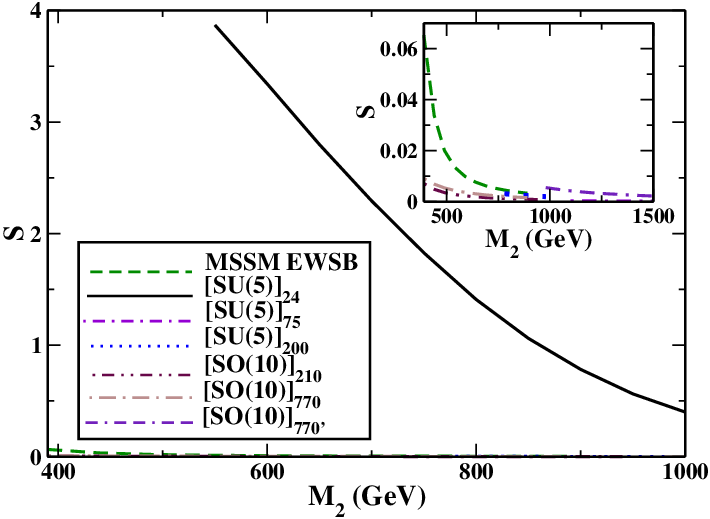}
\caption{The theoretical significance 
$S$ for the radiative neutralino production as a function of the gaugino 
mass parameter  $M_2$ for the different
models with  $\sqrt{s}= 500$~GeV.}
\label{fig:sig_M2}
\end{minipage}
\end{figure}

Finally we address the issue of whether 
the photons coming from the signal process can be measured 
over the photons coming from the background SM process.
The excess of signal photons $N_{\mathrm{S}}=\sigma {\mathcal L}$
over the SM background photons $N_{\rm B}=\sigma_{\rm B}{\mathcal L}$ for a 
given integrated
luminosity $\mathcal{L}$  can be expressed in terms of
the theoretical significance~\cite{Dreiner:2006sb} 
\begin{equation}
S  =  \frac{N_{\rm S}}{\sqrt{N_{\rm S} + N_{\rm B}}}=
\frac{\sigma}{\sqrt{\sigma + \sigma_{\rm B}}} \sqrt{\mathcal L}.
\label{significance}
\end{equation}
\noindent
A theoretical significance of $S = 1$ means that the signal can be measured
at a $68~\%$ confidence level, whereas one needs a significance of $5$  
for the detection of the signal.  
Both the signal and the background process depends significantly on the beam
energy for $\sqrt{s}= 500$~GeV
and $\mathcal L = 5 \times 10^2 $ fb$^{-1}$. In Fig.~\ref{fig:sig_mu}
we show the $\mu$ dependence of the theoretical significance $S$ for the 
different models considered here. When
the lightest neutralino is dominantly a bino type state, then for $\mu$ 
in the range $\mu$ {\Large{$\epsilon$}} [120,160] GeV,
the significance can be a maximum of about  2, 
for the given luminosity. Then the 
signal for the radiative neutralino production would be seen at ILC 
as shown in Fig.~\ref{fig:sig_mu}. 
But for the scenarios with the higgsino as the dominant component
of the lightest neutralino,
it will be  difficult to observe the signal.  
%____________________________________

%________________________
We have also studied the variation of theoretical significance $S$ 
as a function of the gaugino mass parameter $M_2$ as well. The $M_2$ dependence of 
$S$ for all the models considered in this paper
is shown in Fig.~\ref{fig:sig_M2} in the interval 
$M_2$ {\Large{$\epsilon$}} [200,1000] GeV.
The behaviour is similar to $\mu$ with the models having a relatively higher
value of $S$ for lower value of $M_2$.
The values of $S$ given here can be considered as a good guideline,
since we do not include a detector
simulation here.  Besides the theoretical significance, 
one must also consider the signal to background ratio $N_S/N_B$ in order 
to judge the reliability of the analysis. Overall the process under study 
for the different scenarios considered here, specially the one with a 
higgsino type neutralino,
will most probably not be useful for extending the SUSY parameter space 
reach of ILC.
However, it may be possible to reduce the SM background  if 
the electron and positron beams are polarized, with right handed 
electrons and left handed positrons.

  The theoretical significance for most of the scenarios thus considered
is too small, making it difficult to test them in the future linear colliders 
through the radiative neutralino production. But these scenarios have a distinctive 
feature which make it possible to test them, by means of a hard photon tag as employed 
for the process (~\ref{radiative}).

\begin{figure}[htb]
\begin{minipage}[b]{0.4\linewidth}
\centering
\vspace*{0.7cm}
\includegraphics[width=6.3cm, height=5cm]{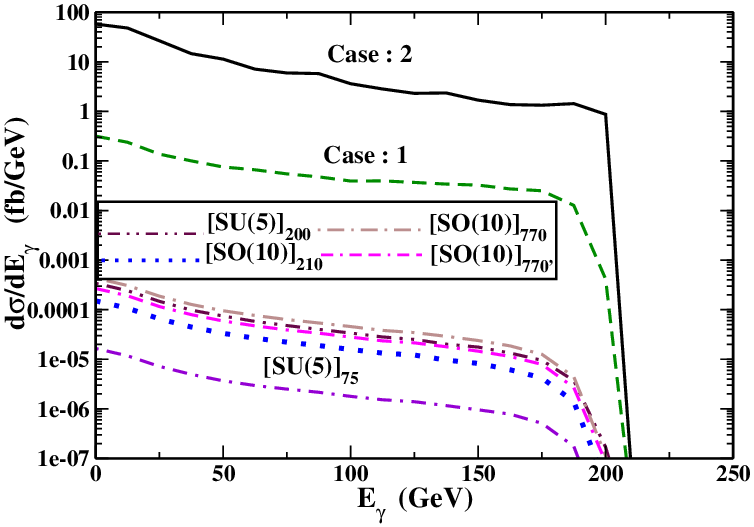}
\caption{The photon energy distribution 
$\frac{d\sigma}{dE_{\gamma}}$ including radiative effects for different cases at $\sqrt{s} = 500$~GeV}
\label{fig:chi_pe}
\end{minipage}
\hspace{0.7cm}
\begin{minipage}[b]{0.4\linewidth}
\centering
\includegraphics[width=6.5cm, height=5cm]{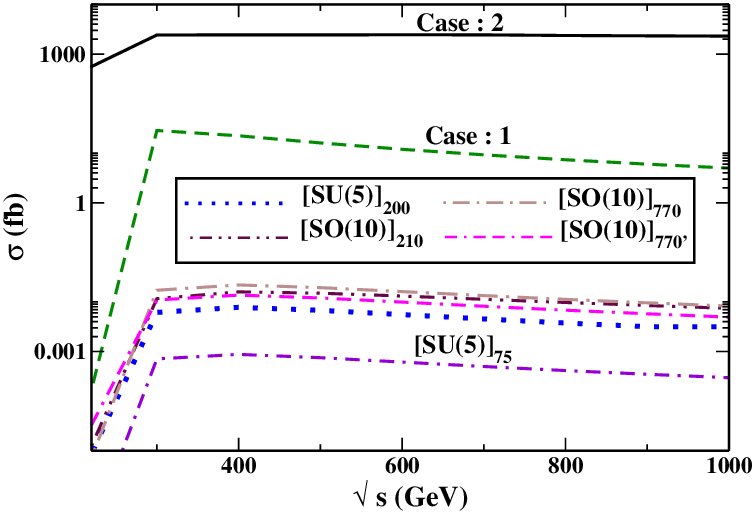}
\caption{The total energy $\sqrt{s}$ dependence for the different cases
defined in the text with the inclusion of radiative effects.}
\label{fig:chi_cs}
\end{minipage}
\end{figure}

For some of the scenarios considered here, namely $[SU(5)]_{75}$, $[SU(5)]_{200}$, $[SO(10)]_{210}$
and $[SO(10)]_{770'}$, due to large values of $M_{1,2}$ and a low value of 
$\mu$, the states ${\tilde {\chi}_1^0},  {\tilde {\chi}_2^0}$
and ${\tilde {\chi}_1^\pm}$ are nearly degenerate,  with mass around $\mu$, and all of them 
having dominant higgsino component. Since they are all 
closely degenerate in
mass, with $m_{\tilde {\chi}_2^0} - m_{\tilde {\chi}_1^0} $ around 15 GeV,
therefore the processes (a)  $e^+e^- \rightarrow {\tilde {\chi}_1^0}{\tilde {\chi}_2^0}\gamma$ and 
(b)  $e^+e^- \rightarrow {\tilde {\chi}_2^0}{\tilde {\chi}_2^0}\gamma$
will also  act as  supersymmetric background to the radiative neutralino production.
The radiative sneutrino production
in these scenarios with a higgsino like LSP has visible decay channels, 
and can, therefore, be easily discriminated.
Since the $Z{\tilde {\chi}_1^0}{\tilde {\chi}_2^0}$ coupling is much larger as compared to  
$Z{\tilde {\chi}_1^0}{\tilde {\chi}_1^0}$ coupling,
therefore a detailed study of the signatures with a hard photon and large missing energy
in the final state will include processes (a), (b) and the radiative neutralino production.
If investigated through the above channel 
the signal for these scenarios will be comparable to the SM irreducible background,
in contrast to the case when only the process (~\ref{radiative}) is considered.
We show in Figs.~\ref{fig:chi_pe} and~\ref{fig:chi_cs}  the photon energy distribution
and the cross section for these models for different cases as follows:

\begin{itemize}
 \item Case 1: \\
Selecting events with $\gamma +$ missing energy in the final state, which
includes the process (a), (b) and the radiative neutralino production.
\item Case 2: \\
The contribution from SM irreducible background, the radiative neutrino production.
\item 
For comparison we have also shown the case of radiative neutralino production ${\tilde {\chi}_1^0} {\tilde {\chi}_1^0} \gamma$
in these Figs.
\end{itemize}

   The radiative effects are included in all these calculations. It can be seen from these Figs. that the selection
of events with $\gamma +$ missing energy in the final state gives almost the same distribution and cross section
for all the models considered. The models can not be discriminated but their  signature is strong compared
to the previous analyses in subsection.~\ref{sec:photon_energ_distribution}. The theoretical significance in this
mode at $\sqrt{s}$ = 500 GeV, increases to about 22, compared to 0.001 for $[SU(5)]_{75}$, for the parameter values given in
Table~\ref{parEWSB75}. Similar result holds for the other scenarios.

%%%%%%%%%%%%%%%%%%%%%%%%%%%%%%%%%%%%%%%%%%%%%%%%%%%%%%%%%%%%%%%%%%%%%%%%%
\section{Summary and Conclusions}
\label{sec:conclusions}
We have carried out a detailed 
analysis of the radiative neutralino production $e^+e^- \to
\tilde\chi^0_1 \tilde\chi^0_1\gamma$  in various GUT models
for the International
Linear Collider energies and compared it with the corresponding 
results in the MSSM with universal gaugino mass parameters. In these models 
the boundary conditions
on the soft gaugino mass parameters can be nonuniversal and 
hence different from those of MSSM with universal boundary conditions.
This process has a signature of a high energy photon and missing energy.
We have  obtained a typical set of parameter values  by excluding certain
regions of the parameter space which follow from 
theoretical and experimental constraints, as discussed in Appendix.~\ref{exp_cons}.

Using this parameter set, we have studied in detail  the 
signal  cross section  for the ILC energies with
unpolarized $e^+$ and $e^-$ beams. For comparison, 
we have used the  MSSM EWSB  scenario as a benchmark.
The contributions from the SM background  $e^+e^- \to \nu \bar\nu \gamma$,
as well as the supersymmetric process $e^+e^- \to \tilde\nu \tilde\nu^\ast \gamma$ acting as a background to
the radiative neutralino production are also considered. All these processes 
have a signature of a highly energetic photon with missing energy.
The photon energy distribution $d\sigma/dE_{\gamma}$, 
and the total cross section as  a function of the total centre of mass energy 
have been calculated for  MSSM EWSB,  
as well as for the different scenarios considered here, 
at $\sqrt{s} = 500$ GeV using  the programme 
CalcHEP.  Since ISR and beamstrahlung are a part of the future linear colliders, due to the planned high
luminosity, for a realistic prediction  we have also included these radiative effects. 
The dependence of the cross section for radiative
neutralino production on the $SU(2)_L$ gaugino mass parameter 
$M_2$ and the Higgs(ino) mass parameter $\mu$, 
as well as its dependence on the selectron~($\tilde e_R, \tilde e_L$)
masses has also been studied and compared  with the corresponding 
results in MSSM EWSB. We have considered scenarios based on 
$SU(5)$ and $SO(10)$ grand unified theories 
with nonuniversal gaugino mass at the grand unified scale.
In the case of $SU(5)$ and $SO(10)$ models we evolve the 
parameters to the electroweak scale and then use these 
to evaluate the radiative neutralino cross section.
All the results mainly depend  on the composition
of the lightest neutralino in the different models considered here.
The models with bino as a dominant component of the lightest
neutralino behave differently
from the models with higgsino as the dominant component of the 
lightest neutralino. The composition of the lightest neutralino depends
on the ratio of the soft gaugino mass parameters at the electroweak scale 
and $\mu$.
The values for these parameters are chosen satisfying all the experimental 
constraints,
along with the requirement that the lightest neutralino is the  LSP.
The supersymmetric background coming from the radiative sneutrino
production is also calculated which depends on the chargino mixing matrix.
The dependence of cross section for these two distinct
scenarios is valid for different  ranges  of the parameters
$M_2$ and $\mu$.
Finally,  in order to understand  whether an excess of signal 
photons, $N_{\mathrm{S}}$, can be measured over the background
photons, $N_{\rm B}$, from radiative neutrino production, we have
analysed the theoretical statistical significance $S = N_{\rm S}/\sqrt{
N_{\rm S} + N_{\rm B}}$, and studied its dependence on the independent 
parameters $M_2$ and 
$\mu$, that enter the neutralino mass matrix.
It is seen that the signal for some scenarios is too weak to be seen at ILC.
Therefore for some of these scenarios which also predict the second lightest
neutralino to be degenerate with the LSP, the observance of 
a signature of a highly energetic photon with missing energy taking into account the second lightest
neutralino production along with LSP production is also considered.
We have also noted that  initial beam polarization
may reduce the background, and
it may be interesting to study whether the signal
for radiative neutralino production  can be enhanced by using
polarized beams. This question in the context of the GUT models will be 
studied in a separate paper~\cite{PNP_Monalisa}.

\section{acknowledgements}
The authors would like to thank B. Ananthanarayan for many
useful discussions.  
P.~N.~P. would like to thank the Centre for High Energy Physics, 
Indian Institute of Science, Bangalore for  hospitality while this
work was initiated. The work of P.~N.~P. is supported by the 
J. C. Bose National Fellowship of the Department of Science and Technology,
and by the Council of Scientific and Industrial Research,
India under the project No.~(03)(1220)/12/EMR-II. 
P.~N.~P would like to thank the Inter-University Centre 
for Astronomy and Astrophysics, Pune, India for hospitality
where part of this work carried out.
%% __________________________________________ 
\appendix
\section{Neutralino Mass Matrix, Lagrangian and Couplings}
\label{neut mass mat}

In this Appendix we summarize the mixing matrix for the neutralinos,
and the couplings that enter our calculations. 
We recall that the neutralino mass matrix receives contribution from 
MSSM superpotential term
\begin{eqnarray}
W_{\mathrm{MSSM}} & = & \mu H_1 H_2,
\label{WMSSM}
\end{eqnarray}
where $H_1$ and $H_2$ are the  Higgs doublet chiral superfields
with opposite hypercharge,
and $\mu$ is the supersymmetric Higgs(ino) parameter. In addition
to (\ref{WMSSM}), the neutralino mass  matrix
receives contributions from the interactions between gauge and  matter
multiplets, as well as contributions from the soft supersymmetry breaking 
masses for the $SU(2)_L$ and $U(1)_Y$  gauginos. 
Putting together all these contributions, the neutralino mass 
matrix, in the bino, wino, higgsino basis 
$(-i\lambda', -i\lambda^3, \psi_{H_1}^1,
\psi_{H_2}^2)$ can be written as~\cite{Bartl:1989ms, Haber:1984rc}
\begin{eqnarray}
\label{mssmneut}
M_{\mathrm{MSSM}} =
\begin{pmatrix}
M_1 & 0   & - m_Z \sw \cos\beta & \phantom{-}m_Z\sw \sin\beta \\
0   & M_2 & \phantom{-} m_Z \cw \cos\beta  & -m_Z \cw\sin\beta \\
 - m_Z \sw \cos\beta &\phantom{-} m_Z \cw \cos\beta  & 0 & -\mu\\
\phantom{-}m_Z\sw \sin\beta& -m_Z \cw\sin\beta & -\mu & 0
\end{pmatrix},
\end{eqnarray}
where $M_1$ and $M_2$ are the $U(1)_Y$ and the $SU(2)_L$
soft supersymmetry breaking gaugino mass parameters, respectively, and
$\tan\beta = v_2 /v_1$ is the ratio of the vacuum expectation
values of the neutral components of the two Higgs doublet 
fields $H_1$ and $H_2$, respectively. Furthermore,
$m_Z$ is the $Z$ boson mass, and $\theta_W$ is the
weak mixing angle. We shall consider all parameters in the neutralino mass
matrix to be real. In this case it 
can be diagonalised by an orthogonal matrix. 
If one of the 
eigenvalues of $M_{\rm MSSM}$ is negative, one can diagonalize
this matrix using a unitary matrix $N$, the neutralino 
mixing matrix, to get a positive semi definite diagonal 
matrix~\cite{Haber:1984rc} 
with the neutralino masses $m_{\chi_i^0}~(i = 1, 2, 3, 4)$ in order of
increasing value: 
\begin{eqnarray}
\label{mssmdiag}
N^\ast M_{\mathrm{MSSM}} N^{-1} =   \mathrm{diag}\begin{pmatrix}m_{\chi_1^0}, 
& m_{\chi_2^0}, & m_{\chi_3^0}, & m_{\chi_4^0} \end{pmatrix}.
\end{eqnarray}
%____________________________________________________________________

For the minimal supersymmetric standard model the 
interaction Lagrangian of neutralinos, electrons, selectrons and 
$Z$ bosons  is given by~\cite{Haber:1984rc}
\begin{eqnarray}
{\mathcal L} &=& (- \frac {\sqrt{2}e}{\cw} N_{11}^*)
                    \bar{f}_eP_L\tilde{\chi}^0_1\tilde{e}_R
                 + \frac{e}{\sqrt{2} \sw} (N_{12} + \tw N_{11}) 
                   \bar{f}_e P_R\tilde{\chi}^0_1\tilde{e}_L \nonumber \\
     & & + \frac{e}{4 \sw \cw} \left(|N_{13}|^2 - |N_{14}|^2\right)
           Z_\mu \bar{\tilde{\chi}}_1^0\gamma^\mu \gamma^5\tilde{\chi}_1^0
           \nonumber \\
          && + e Z_\mu \bar{f}_e \gamma^\mu 
             \big[ \frac{1}{\sw\cw}\left(\frac{1}{2} - \sw[2]\right) P_L 
                     - \tw  P_R\big] {f}_e + \mathrm{h. c.},
\label{mssmlagrangian}
\end{eqnarray} 
with the electron, selectrons, neutralino and $Z$ boson fields denoted by
$f_e$, $\tilde{e}_{L,R}$, $\tilde{\chi}_1^0$, and $Z_\mu$, respectively,  
and $P_{R, L} = \frac{1}{2} \left(1 \pm \gamma^5\right)$. 
The interaction vertices 
following from (\ref{mssmlagrangian}) are summarized 
in Table~\ref{feynmandiag}.

%%%%%%%%%%%%%%%%%%%%%%%%%%%%%%%%%%%%%%%%%%%%%%%%%%%%%%%%%%%%%%%%%%%%%%%%%%
%%%%%%% Vertices of MSSM for Radiative Neutralino Production %%%
%%%%%%%%%%%%%%%%%%%%%%%%%%%%%%%%%%%%%%%%%%%%%%%%%%%%%%%%%%%%%%%%%%%%%%%%%%
\begin{table}[h!]
\begin{center}
\caption{Vertices corresponding to different terms in the interaction 
Lagrangian (\ref{mssmlagrangian}) for MSSM. Here we have 
also shown the vertices for selectron-photon  and electron-photon
interactions~\cite{Basu:2007ys}.}
\vspace{5mm}
\begin{tabular}{lccccccccl}
\hline
\\
Vertex & & & & & & & & & Vertex Factor\\
& & &  & &&& &&\\
\hline
& & & & & && &&\\
%%%%%%%%%%%%%%%%%%
right~selectron - electron - neutralino 
& & & & & && && {$\frac {-i e \sqrt{2}}{\cw}N_{11}^* P_L$}\\
& & & & & && &&\\
%%%%%%%%%%%%%%%%%%%%
left~selectron - electron - neutralino
& & & & & && &&{$\frac{i e}{\sqrt{2} \sw} (N_{12} + \tw N_{11}) P_R$}\\
& & & & & && &&\\
%%%%%%%%%%%%%%%%%%%%%%%
neutralino - $Z^0$ - neutralino
& & & & & && && {$ \frac{i e}{4 \sw \cw} 
           \left(|N_{13}|^2 - |N_{14}|^2\right) \gamma^\mu \gamma^5$}\\
& & & & & && && \\
%%%%%%%%%%%%%%%%%%%%%%%%%%%
electron - $Z^0$ - electron
& & & & & && && {$ i e \gamma^\mu \big[ \frac{1}{\sw\cw}\left(\frac{1}{2} 
               - \sw[2]\right) P_L - \tw  P_R\big] $} \\
& & & & & && &&\\
%%%%%%%%%%%%%%%%%%%%%%.
selectron - photon - selectron
& & & & & && && {$ i e (p_1 + p_2)^\mu$}\\
& & & & & && && \\
%%%%%%%%%%%%%%%%%%%%%%%%%%%%%%%
electron - photon - electron
& & & & & && && {$ i e \gamma^\mu$}\\
& & & & &  && && \\
%%%%%%%%%%%%%%%%%%%%%%%%%%%
\\
\hline
\bottomrule
\end{tabular}
\label{feynmandiag}
\end{center}
\end{table}
%%%%%%%%%%%%%%%%%%%%%%%%%%%%%%%%%%%%%%%%%%%%%%%%%%%%%%%%%%%%%%%%%%%%%%%
%%%% End of Vertices of MSSM   for Radiative Neutralino Production %%%
%%%%%%%%%%%%%%%%%%%%%%%%%%%%%%%%%%%%%%%%%%%%%%%%%%%%%%%%%%%%%%%%%%%%%%%%%
\section{ Experimental Constraints}
\label{exp_cons}
In this Appendix we review the constraints on the SUSY particle
spectrum from the different experiments at LHC, Tevatron and LEP.

\subsection{Limits on Gaugino mass parameters}

The exclusion limit on the gaugino mass parameters is set 
from the current experimental limits on the 
superpartner masses. Since no supersymmetric partners of the SM
particles have been detected in the experiments, only lower
limits on their masses have been obtained.  In particular, the
search for the lightest chargino state at LEP has yielded lower
limits on its mass \cite{lep-chargino}.
The lower limit depends on the spectrum of the model \cite{Yao:2006px}.
Assuming that $m_0$, the soft supersymmetry breaking scalar mass,  is large, 
the  limit on the lightest chargino mass  following from 
non observation of chargino pair production in $e^+ e^-$
collisions is 
\be 
M_{\tilde \chi_1^{\pm}} \gsim 103~~{\rm GeV}.  \label{ch-limit}
\ee 
The limit depends on the sneutrino mass.  For a sneutrino mass below 200
GeV, the bound becomes weaker, since the production of a chargino pair
becomes more rare due to the destructive interference between $\gamma$ or
$Z$ in the $s$-channel and $\tilde\nu$ in the $t$-channel. In the models we
consider, $m_{\tilde\nu}$ is close to $m_0$.  When $m_{\tilde\nu}<200$
GeV, but $m_{\tilde\nu}>m_{\tilde\chi^\pm}$, the limit 
becomes \cite{Yao:2006px}

\be M_{\tilde \chi_1^{\pm}} \gsim 85~~{\rm GeV}.  \ee
For the parameters of the chargino mass matrix, the limit
(\ref{ch-limit}) implies 
an approximate lower limit~\cite{Abdallah:2003xe,Dreiner:2009ic} 
\be M_2,~~ \mu \gsim 100~~{\rm GeV}.
\label{limits1}
\ee 
The limits shown in Eq.~(\ref{limits1}) on  $M_2$ and $\mu$ are 
obtained by scanning over the MSSM parameter space and are, therefore,
expected to be model independent~\cite{Huitu:2010me}. 
From the chargino mass limit from LEP, a neutralino mass limit is obtained
assuming gaugino mass unification at
high energy scales and amounts to 47 GeV.  But due to the strong constraints
from LHC experiments, the lower limit on $\tilde \chi_1^0$ in case of constrained MSSM (CMSSM)
has risen to about 100 GeV.

\subsection{Exclusion limits on squarks and gluinos}

Experiments at the LHC and Tevatron being the proton-(anti) proton collider, with their higher
centre of mass energies compared to LEP, is more sensitive for the SUSY particles
carrying color charge, squarks and gluinos, because of QCD-mediated processes. Limits of the order
of about 100 GeV, was set on the squark masses by LEP, but the recent hadron collider
experiments have set much higher limits.

   Gluino masses below 800 GeV, are excluded by both ATLAS and CMS collaboration
in the framework of CMSSM for all squark masses. For equal squark and gluino masses,
the limit is around 1400 GeV~\cite{Atlas_gluino, CMS_gluino}. These results are only slightly dependent on the choice
of the CMSSM parameters, $\tan \beta, A_0$ and $\mu$. Similar analyses has been carried
out in the context of simplified models, where upper limits on gluino pair production
are derived as a function of the gluino and the lightest neutralino mass. For massless LSP,
$m_{\tilde g} <$ 900 GeV is excluded. For a heavy LSP above 300 GeV, no general limit on
gluino can be  set.

    Limits on the first two generation squark masses are set by the LHC experiments, with
lowers limits of around 1300 GeV~\cite{Atlas_gluino, CMS_gluino}, for all gluino masses in the framework of CMSSM. The  analyses for squarks 
are similar to the gluinos. For massless neutralino in the framework of simplified models,
squark masses below 750 GeV are excluded, whereas increasing the mass of LSP above 200 GeV leads to a 
degradation of the limits.

   The limits on the third generation $\tilde t_1$ mass from LEP was around 96 GeV, in
the charm plus neutralino final state~\cite{lep-chargino}. LHC along with Tevatron have performed the analyses 
for third generation squarks in different
scenarios, leading to different final states~\cite{cms-stop}. Similar analyses has been carried out for the sbottom
quarks. Overall, for our analyses we are considering the scenario where the third generation squarks are excluded 
below a mass of about  800 GeV.

 \subsection{Exclusion limit on slepton masses} 
 
  The strongest limit on the slepton masses come from LEP, because of
its clean signature. LEP experiments~\cite{lep-chargino} have excluded sleptons for masses below 100 GeV in case
of different scenarios.  Similarly for the sneutrinos, limits are obtained from
the invisible width of the $Z$ boson along with the limits derived from the searches of
gauginos and sleptons. They are excluded for a mass of about  94 GeV.

    Taking into account all the above constraints set by the LEP and LHC experiments, for
our analyses we have taken $m_{\tilde g} \approx$ 1400 GeV, mass of first two generation squarks
around 1300 GeV, $m_{\tilde t}$ around 1000 GeV along with the slepton
masses around 150 GeV.

%_________________________________________________________
%\newpage
%__________________________________________________________

%%%%%%%%
\end{document}